\documentclass[12pt]{article} 
\def\cdate{{December 5, 2024}}
\usepackage{amsfonts,amsmath,latexsym,amssymb,
txfonts} 
\usepackage{amscd}
\usepackage[american]{babel}
\makeindex
\usepackage{color}
\usepackage{graphicx}
\makeatletter
\renewcommand{\@evenhead}{\raisebox{0pt}[\headheight][0pt]{\vbox{\hbox
to \textwidth{\thepage\hfil\strut\textsc{\leftmark}}\hrule}}}
\renewcommand{\@oddhead}{\raisebox{0pt}[\headheight][0pt]{\vbox{\hbox
to \textwidth{\textsc{\rightmark}\hfil\strut\thepage}\hrule}}}
\makeatother
\makeatletter
\def\timenow{
\@tempcnta=\time \divide\@tempcnta by 60 \number\@tempcnta:\multiply
\@tempcnta by 60 \@tempcntb=\time \advance\@tempcntb by -\@tempcnta
\ifnum\@tempcntb <10 0\number\@tempcntb\else\number\@tempcntb\fi}
\newcounter{outputpage}
\renewcommand{\@oddhead}
{\stepcounter{outputpage}\hfill\hfill\theoutputpage}
\renewcommand{\@evenhead}
{\stepcounter{outputpage}\hfill\hfill\theoutputpage}
\renewcommand{\@oddfoot}
{\vbox{
\hrule
\vspace{3pt}
\hfil
{\scriptsize\textit{
\hfill\hfill\jobname.tex; \today; \timenow; p. \theoutputpage}}
\hfil
}}
\renewcommand{\@evenfoot}
{\vbox{
\hrule
\vspace{3pt}
\hfil
{\scriptsize\textit{
\hfill\hfill\jobname.tex; \today; \timenow; p. \theoutputpage
}}
\hfil
}}
\makeatother
\def\RR{{\mathbb R}} 
\def\CC{{\mathbb C}} 
\def\HH{{\mathbb H}}

\def\PP{{\mathbb P}}

\def\cA{\mathcal{A}}

\def\cD{\mathcal{D}}

\def\cF{\mathcal{F}}

\def\cM{\mathcal{M}}

\def\cR{\mathcal{R}}
\def\cV{\mathcal{V}}

\def\cH{\mathcal{H}}
\def\cG{\mathcal{G}}

\def\cX{\mathcal{X}}
\def\cY{\mathcal{Y}}
\def\cZ{\mathcal{Z}}

\def\Tr{\mathrm{ Tr\,}}

\def\vol{\mathrm{ vol\,}}

\def\la{\langle}
\def\ra{\rangle}

\def\nn{\nonumber} 
\def\be{\begin{equation}} 
\def\ee{\end{equation}} 
\def\bea{\begin{eqnarray}} 
\def\eea{\end{eqnarray}} 
\def\bed{\begin{definition}{\ }}
\def\eed{\end{definition}}
\def\ed{\end{document}}
\def\bp{\begin{proposition}}
\def\ep{\end{proposition}}
\def\bc{\begin{center}}
\def\ec{\end{center}}
\def\bi{\begin{itemize}} 
\def\ei{\end{itemize}} 
\def\benum{\begin{enumerate}} 
\def\eenum{\end{enumerate}} 
\def\bmp{\begin{minipage}} 
\def\emp{\end{minipage}}

\newtheorem{proposition}{Proposition}

\newtheorem{definition}{Definition}

\begin{document}

\begin{titlepage}
\thispagestyle{empty}
\null
\vspace{-3cm}
\hspace*{50truemm}{\hrulefill}\par\vskip-4truemm\par
\hspace*{50truemm}{\hrulefill}\par\vskip5mm\par
\hspace*{50truemm}{{\large\sc 
New Mexico Tech {\rm 
(\cdate)
}}}\vskip4mm\par
\hspace*{50truemm}{\hrulefill}\par\vskip-4truemm\par
\hspace*{50truemm}{\hrulefill}
\par
\bigskip
\bigskip
%\draft
\par
%\hspace*{50truemm}{\LARGE\textbf{\textsf{DRAFT}}}
\par
\vspace{1cm}
\centerline{\huge\bf Geometric Deformation}
\bigskip
\centerline{\huge\bf of Quantum Mechanics}
%\bigskip
\bigskip
\bigskip
\bigskip
\centerline{\Large\bf Ivan G. Avramidi and Roberto Niardi}
\bigskip
\centerline{\it Department of Mathematics}
\centerline{\it New Mexico Institute of Mining and Technology}
\centerline{\it Socorro, NM 87801, USA}
\centerline{\it E-mail: ivan.avramidi@nmt.edu, 
roberto.niardi@student.nmt.edu}
\bigskip
%\centerline{Revised on }
%\medskip
%\maketitle 
\vfill
{\narrower
\par
% Abstract

We develop a novel approach to Quantum Mechanics that we call Curved Quantum Mechanics.
We introduce an infinite-dimensional 
K\"ahler manifold $\cM$, that we call the state manifold, such that the 
cotangent space $T_z^*\cM$ is a Hilbert space.
In this approach a state of a quantum system is described by
a point in the cotangent bundle $T^*\cM$, 
 that is, by a point $z\in\cM$ in the state manifold and a one-form $\psi\in T^*_z\cM$.
The quantum dynamics is described by an infinite-dimensional Hamiltonian system 
on the state manifold with a magnetic field $H$, which reduces to the Schr\"odinger equation
for zero curvature and reduces to the equations of geodesics for zero magnetic field.
The curvature of the state manifold is determined by gravity, that is, by
the mass/energy of the system,
so that for microscopic systems the manifold is flat and for
macroscopic systems it is strongly curved, which prohibits Schr\"odinger cat
type states. We solved the dynamical equations exactly for the complex projective space
and the complex hyperbolic space and show that in the case of negative curvature
there is a bifurcation at a critical value of the curvature.
This means that for small mass all modes are in the quantum regime with the unitary 
periodic dynamics and for large mass there are classical modes, with not a periodic
but rather an exponential time evolution leading to a collapse of the state vector.

\par}

\vfill

\end{titlepage}

%\pagenumbering{Roman}
%\tableofcontents
%\markboth{Contents}{Contents}

\section{Introduction}
\setcounter{equation}{0} 

Quantum Mechanics, with its probabilistic nature and wave-particle duality, has
been the cornerstone of modern physics for over a century. A fundamental
principle of Quantum Mechanics is the superposition principle requiting a linear
structure for its adequate description. The mathematical framework suited for
such a description is the theory of complex Hilbert spaces and the corresponding
spectral theory of self-adjoint operators. It has provided remarkable predictive
power and theoretical insights into the microscopic world.

However, quantum mechanics in its currect state has foundational limitations,
particularly when trying to reconcile it with general relativity or explain
phenomena at very small scales, which motivates the search for modifications.
Some modifications focus on reinterpreting the meaning of the quantum state,
like the ``many-worlds interpretation'' 
\cite{everett73} or ``Bohmian Mechanics'' \cite{bohm52a,bohm52b}, without 
necessarily altering the core equations. Some other models
introduce additional variables, like ``hidden variables'', to potentially
explain seemingly random quantum behavior. This could involve adjusting the
basic equations of quantum mechanics by introducing non-linear terms. The so
called collapse models propose a mechanism for wavefunction collapse that occurs
spontaneously, potentially explaining the transition from quantum to classical
behavior. In particular, one is argued
that gravity might be responsible for the
collapse of the wavefunction 
\cite{penrose96,penrose14}. The most well-known examples of such
theories are: Ghirardi–Rimini–Weber model 
\cite{ghirardi86}, continuous spontaneous localization model 
\cite{ghirardi90} and Di\'osi–Penrose model \cite{penrose96, diosi89}.

Nevertheless, researchers have been increasingly intrigued by the possibility of
a nonlinear extension of Quantum Mechanics
\cite{weinberg89,castro06,penrose14,perelman20}. This endeavor seeks to
reconcile the principles of quantum theory with the curved structures observed
in General Relativity, aiming to construct a unified framework that can
encompass both quantum and gravitational phenomena.

One compelling approach in this pursuit is the exploration of nonlinear
generalizations of quantum mechanics on curved Hilbert manifolds. By allowing
the Hilbert space to possess a nontrivial curvature, this novel perspective
transcends the limitations imposed by traditional quantum theory. Such an
extension holds the promise of providing a deeper understanding of the interplay
between geometry and quantum phenomena, potentially unlocking new insights into
the fundamental nature of our universe.

The study of quantum mechanics on curved Hilbert manifolds not only addresses
theoretical gaps but also has profound implications for a broad range of
scientific disciplines. It has the potential to shed light on the behavior of
quantum systems in the vicinity of black holes, where strong gravitational
fields significantly deform the surrounding spacetime. Additionally, it may
offer fresh perspectives on the early Universe, where quantum effects played a
pivotal role during cosmic inflation.

This paper is an attempt of generalizing the standard Quantum Mechanics. We aim
to explore the theoretical foundations underlying this framework, examining the
mathematical formalism that enables the description of quantum phenomena in
curved spaces. Furthermore, we will delve into the potential implications and
applications of this nonlinear extension, envisioning its impact on various
branches of physics, including quantum gravity and cosmology.

Through a systematic analysis of existing research and theoretical developments,
we aim to contribute to the growing understanding of the interplay between
quantum mechanics and curved spacetime geometries. By bridging the gap between
quantum theory and General Relativity, we strive to pave the way for a more
unified description of the fundamental laws governing our universe.

Overall, this paper sets out to present an exploration of the nonlinear
generalization of quantum mechanics on curved Hilbert manifolds, aiming to
provide researchers with a deeper understanding of this fascinating field and
its potential implications for the future of physics.

%\section{Introduction} %\setcounter{equation}0
Our main idea is to replace the flat Hilbert vector space by a curved Hilbert
manifold whose tangent spaces are Hilbert spaces. In this picture, the states of
a system are not the vectors of a Hilbert space but points in the cotangent
bundle of the Hilbert manifold. Therefore, one cannot just add any two states.
That will certainly put restrictions on the superposition principle. To add
states one needs some kind of parallel transport on the manifold which is
supposed to be defined by gravity.

The idea is then that the curvature of this infinite-dimensional manifold is
determined by gravity, that is, by the mass (or the energy) of the system. That
is why, the true quantum phenomena are observable only for microscopic systems
and are washed out for heavy macroscopic systems. Of course, depending on the
system the state manifold could have some flat directions, which means that some
states can always be added.

It is easy to see that the standard linear Schr\"odinger equation can be viewed
as a rather trivial infinite-dimensional Hamiltonian system on the cotangent
bundle to a Hilbert manifold. In our picture this Hamiltonian system is more
general with some nontrivial metric and connection. The quantum dynamics of
large macroscopic systems should converge to classical dynamics, including the
gravitational phenomena, without paradoxes like the Scr\"odinger cat.

It is worth noting that our approach is geometric in nature and is radically
different from the Schr\"odinger-Newton equation \cite{diosi84,penrose96} in 
$\RR^3$, which has
the form
\be
\frac{\partial\psi_t(x)}{\partial t}
=-\frac{i}{\hbar}\left(
-\frac{\hbar^2}{2m}\Delta
+V(x)\right)\psi_t(x)
+i\frac{Gm^2}{\hbar}
\int\limits_{M} dy\;\frac{1}{|x-y|}\psi_t(y)\bar\psi_t(y)
\psi_t(x).
\ee
Even though this equation is non-linear, the type of the non-linearity is different:
it is cubic in $\psi$ rather than quadratic as in our approach.

%===================================================================
%EDIT THIS!!!!!!!!!!!!!!!!!!

%===================================================================

%\section{Standard Quantum Mechanics}
% \setcounter{equation}{0} 
%\section{Standard Quantum Mechanics}

To better explain our main idea let us recall first the standard formulation
of Quantum Mechanics.
The mathematical foundation of quantum mechanics is the spectral theory of
self-adjoint operators on Hilbert spaces. %\begin{enumerate}
%\item
A Hilbert space $\cV$ is a complex vector space equipped with an inner product
$(\cdot,\cdot)$ and the corresponding norm
$||\cdot||$. Self-adjoint operators are linear maps on the Hilbert space with real
spectrum.
The main postulates of the quantum mechanics are:

\begin{enumerate} 

\item A quantum physical system is described in terms of two sets of objects:
{\it observables} and {\it states}.

\item A system can be either in {\it pure} or a {\it mixed} state. A pure state is a nonzero
vector $\psi\in \cV$ in a complex Hilbert space $\cV$. A mixed state is a
self-adjoint positive operator $\rho: \cV\to \cV$ on the Hilbert space $\cV$ of
trace one, called the {\it density operator}. We will only consider pure states in
this paper, if not specified otherwise.

\item Two vectors, $\psi$ and $\psi'=\alpha\psi$, with any non-zero complex number
$\alpha\in \CC$, describe the same state. Similarly, two density operators,
$\rho$, and $\rho'=U\rho U^{-1}$, with a unitary operator $U$, describe the same
state. That is, the set of states is, in general, the { infinite-dimensional
complex projective space} ${\CC \PP}^\infty$.

\item For any two states $\psi_1$ and $\psi_2$ any linear combination
$\psi'=\alpha_1\psi_1+\alpha_2\psi_2$, with $\alpha_1,\alpha_2 \in\CC$, defines
a physical state of the system.

%==============================
%\item
%The set of observables is a { real algebra}.

\item
Every physical observable is described by a self-adjoint
operator 
\be
A: \cV\to \cV
\ee
on a Hilbert space
with a real spectrum $\sigma(A)\subseteq \RR$.

\item
The set of all possible outcomes of a measurement of an observable $A$ 
is a subset of the spectrum of the operator $A$,
that is, the results of a measurement can only be
its eigenvalues.

\item
The {\it measurement} of an observable $A$ in a pure state $\psi$ is a 
{\it random variable} (usually called the {\it collapse
of the wave function})
\be
X_A: \cV\to \sigma(A).
\ee
The random variable $X_A$ maps an eigenvector $\psi_\lambda$
of the operator $A$ with an eigenvalue $\lambda$ to the eigenvalue $\lambda$,
\be
X_A(\psi_\lambda)=\lambda.
\ee
This random variable induces the probability distribution
\be
\mu_A(\lambda,\psi)=\frac{|(\psi_\lambda,\psi)|^2}{||\psi_\lambda||^2\;||\psi||^2},
\ee
where $\lambda=X_A(\psi)$ is an eigenvalue of the operator $A$ and
$\psi_\lambda$ is the corresponding eigenvector. In other words,
$\mu_A(\lambda,\psi)$ is the probability of the result of the measurement of the
observable $A$ in a state $\psi$ being equal to the eigenvalue $\lambda$ of the
operator $A$. This is usually called the {\it Born rule}.

\item
The {\it expectation value} of the measurement of an observable $A$
in a state $\psi$ is
\be
\left<A\right>_\psi=\frac{(\psi,A\psi)}{||\psi||^2}
=\sum_{\lambda\in\sigma(A)}\mu_A(\lambda,\psi)\lambda.
\ee
Similarly,
\be
\la A\ra_\rho=\Tr (\rho A).
\ee

\item
The {\it dynamics} of the quantum system is determined by 
 a self-adjoint operator bounded from below 
(called the {\it Hamiltonian})
\be
H: \cV\to \cV,
\ee
which generates the spectral decomposition of the Hilbert space
\be
\cV=\bigoplus_{k=1}^\infty V_k,
\label{221iga}
\ee
where $V_k$ are the finite-dimensional complex vector spaces,
which are the eigenspaces of the Hamiltonian
with eigenvalues $H_k$, so that the Hamiltonian is just a multiplication
operator by $H_k$ on each subspace $V_k$.

\item
The quantum dynamics is governed by
the Schr\"odinger equation
\be
\frac{d\psi}{dt} = -\frac{i}{\hbar}H\psi,
\ee
or
\be
\frac{d\rho}{dt} = -\frac{i}{\hbar}[H,\rho],
\ee
where $\hbar$ is the Planck constant.
This leads to the {\it unitary time evolution}
\be
\psi(t)=U(t)\xi
\ee
or
\be
\rho(t)=U(t)\rho_0U(-t),
\ee
where $\xi{}=\psi(0)$, $\rho(t)=\rho_0$, and
\be
U(t)=\exp\left(-\frac{i}{\hbar}tH\right),
\ee
is a unitary operator,
so that
\be
||\psi(t)||=||\xi{}||.
\ee

\item
The unitary dynamics of Quantum Mechanics via the Schr\"odinger equation 
%(\ref{})
can be described by an infinite-dimensional
{\it Hamiltonian system} $(z,\varphi)$
with a quadratic Hamiltonian
\be
\cH(z,\bar z,\varphi,\bar\varphi)
=\left(\psi,\psi\right),
%=||D\psi||^2,
\ee 
where 
\be
\psi=\varphi-\frac{i}{2\hbar}H\bar z.
\ee
%and $H$ is the Hamiltonian.
Then the Hamiltonian system is reduced to
\bea
\frac{dz}{dt} &=& \bar\psi,
\\
\frac{d\psi}{dt} &=& -\frac{i}{\hbar}H\psi.
\eea
Of course, these two equations decouple; the
dynamical trajectory with the initial conditions
\be
z(0)=0
\qquad
\psi(0)=\xi{},
\ee
is
\be
z(t) = \bar Z(t)\bar\xi,
\qquad
\psi(t) = U(t)\xi{},
\ee
where
\bea
%U(t) &=& \exp\left(-\frac{i}{\hbar}tH\right),
%\\
Z(t) &=& \int_0^t d\tau\;  U(\tau).
%\nn\\
%&=&
%i\hbar \; {H^{-1}}\left\{U(t)-1\right\}.
\eea
%So, $\psi(t)$ is completely independent on $z(t)$.
%In the following we set $\hbar=1$.

\end{enumerate}

%=======================================================

%\section{Curved Quantum Mechanics}
%\setcounter{equation}{0} 

%=================================
%==================================================
%\section{Curved Quantum Mechanics}
%\setcounter{equation}{0} 

We generalize this picture in a geometric language.
We use the analogy with General Relativity. Einstein 
Relativity Principle (in Special Relativity)
postulates that the laws of all physics (except gravity) are identical in all
inertial reference frames, that is, they are invariant under the Lorentz
transformations, which are linear pseudo-Euclidean transformations of the
coordinates of the Minkowski spacetime with a flat pseudo-Euclidean metric.
Gravity is then described in General Relativity by the curvature of a
pseudo-Riemannian metric in a curved spacetime. It is postulated that the laws
of physics are now identical in all (not just inertial) reference frames, that
is, {\it the theory is invariant under all smooth diffemorphisms} (non-linear coordinate
transformations) of spacetime.
We propose to generalize the postulates of quantum mechanics as follows:

\begin{enumerate}
\item
A quantum physical system is described by an
{\it infinite-dimensional almost complex manifold} $\cM$ (that we call the 
{\it state manifold})
with a Hermitian metric and the corresponding
inner product $(\;,\;)$.
More specifically, we will assume that the state manifold
is an {\it infinite-dimensional homogeneous Einstein-K\"ahler manifold}
such as the infinite-dimensional complex projective space $\CC\PP^\infty$
and the infinite-dimensional complex hyperbolic space $\CC\HH^\infty$.
The cotangent space $T^*_z\cM$ of this manifold at every point $z$ is the
familiar complex Hilbert space.

\item We introduce a {\it Quantum Relativity Principle} as follows. A
holomorphic coordinate chart $(z,\bar z)$ in this manifold describes an {\it observer}
(an analogy with the reference frame). The standard Quantum Mechanics (without
gravity) is {\it observer invariant}, that is, invariant under the linear
transformations of the coordinates of the flat Hilbert space (similar to Special
Relativity).

\item {\it Gravity is described by the curvature of a curved metric on the state
manifold $\cM$, which is determined by the mass/energy of the system}. It is postulated
that {\it the theory is invariant under the group of holomorphic diffeomorphisms of
this manifold} (similar to General Relativity).

%\item
%It is assumed that there is a neighborhood $\cQ$
%of the origin such that every point in this neighborhood can be 
%connected to the
%origin by a unique geodesic. We call it the
%entanglement (or quantum) region.

\item We interpret the cotangent bundle $T^*\cM$ as follows: {\it the base state
manifold $\cM$ describes the environment} whereas {\it the cotangent space 
$T_z^*\cM$ describes
the quantum system}. A state of the whole system is described by a point in the
cotangent bundle $T^*\cM$, that is, by a one-form $\psi\in T^*_z\cM$ at a point
$z\in\cM$. For any real number $\alpha\in\RR$ the covectors $\psi$ and
$\psi'=\exp(i\alpha) \psi$ describe the same state.

%That is, a state is a point in a vector bundle with the
%typical fiber being an infinite dimensional complex projective space,
%so, locally it is the product $\cM\times \CC\PP^\infty$.

%\item
%Every point $z(t)$ on the geodesics connecting it to the origin
%describes the same state. 

\item The standard superposition principle of quantum mechanics is restricted as
follows. In general, {\it one cannot add arbitrary states to form new states} since
states are covectors in {\it different} cotangent spaces. Let $\psi_1\in
T^*_{z_1}\cM$ and $\psi_2\in T^*_{z_2}\cM$ be two states at the points $z_1$ and
$z_2$.

\benum 
\item If the points $z_1$ and $z_2$ are sufficiently close to each other
then there is a unique geodesic connecting these points. Then one can parallel
transport the covector $\psi_2$ along the geodesic from the point $z_2$ to the
point $z_1$ to form a state $\tilde \psi_2$ at the point $z_1$ and define a new
state by the superposition of these two covectors, $\psi'=\alpha_1
\psi_1+\alpha_2\tilde \psi_2$.

\item If the points $z_1$ and $z_2$ cannot be connected by a unique geodesic
then there is no way to form a new state by a superposition of the states
$\psi_1$ and $\psi_2$.

\item { The curvature of the state manifold is be determined by the gravitational
mass of the system, so that for microscopic systems the manifold is flat and for
macroscopic systems it is strongly curved.} This prohibits the Schr\"odinger cat
type states.

\eenum

\item Every physical observable is described by a Hermitian tensor of type
$(1,1)$ that can be identified with a self-adjoint operator on the cotangent
space 
\be A: T^*_{z}\cM\to T^*_{z}\cM. 
\ee

%such that for any $\psi,\varphi\in T_{z}\cM$.
%Note that if the map $A$ is linear then $A(z)=A^*z$.

%\item
%The probability of the result $\lambda$ of the measurement of the
%observable $A$ in a general state $z$ is
%\be
%\frac{|h(\psi_\lambda,\psi)|^2}{h(\psi,\psi)}
%\ee

\item
The expectation value of the measurement of the observable $A$
in a state $(z,\psi)$ is
\be
\left<A\right>_\psi=\frac{(\psi,A\psi)}{||\psi||^2}.
\ee
%where $(\psi,\varphi)$ denotes the inner product in the cotangent
%space $T^*_z\cM$ defined with the metric of the K\"ahler manifold
%and $||\psi||^2=(\psi,\psi)$.

%\item
%The dynamics of the system is described by the equation
%\be
%\frac{d}{dt} z=
%\ee

%\item 
Under the holomorphic change of the coordinates the state of the system,
$\psi$, transforms as a one-form and the observable, $A$, transforms as a
$(1,1)$ tensor, so that the expectation value $\la A\ra_\psi$ is invariant.

\item
The Hamiltonian of standard quantum mechanics is replaced by an operator
\be
H: T_z^*\cM\to T^*_z\cM,
\ee
acting on the cotangent space of the state manifold.
We replace the spectral decomposition of the standard quantum mechanics
(\ref{221iga}) by the factorization of the state manifold.
More precisely, each eigenspace $V_k$, which is a finite-dimensional
complex vector space, is replaced with a 
{\it curved K\"ahler-Einstein manifold}, $M_k$,
so that the whole state manifold is
\be
\cM=\varprod_{k=1}^\infty M_k,
\ee
and the cotangent spaces decompose accordingly
\be
T_z^*\cM=\bigoplus_{k=1}^\infty T_z^*M_k,
\ee
where $T_z^*M$ are finite-dimensional complex vector spaces.
The Hamiltonian operator $H$ is constant on each finite-dimensional 
cotangent space $T_z^*M_k$; it is just the operator of multiplication by a real constant
$H_k$. That is, we propose to {\it curve the eigenspaces of the Hamiltonian, with
the curvature being determined by gravity}.

\item The dynamics of the quantum system is described by an {\it infinite-dimensional
Hamiltonian system} on the state manifold such that in the limit of small mass it
reduces to the Schr\"odinger equation and in the limit of large mass it reduces
to the equation of geodesic on the state manifold. The infinite-dimensional
curved state manifold $\cM$ can have finite-dimensional submanifolds (which can
be flat, in particular).

\end{enumerate}

This paper is organized as follows.
In Sec. 2 we describe the geometry of K\"ahler manifolds,
both finite-dimensional and infinite-dimensional.
In particular, we explore the geometry of the complex
projective spaces and the complex hyperbolic spaces.
In Sec. 3  describe the Hamiltonian systems on K\"ahler manifolds with an
Einstein-K\"ahler magnetic field. In the case of the complex
projective space and the complex hyperbolic space we 
solve the dynamical equations exactly. We show that in the case of the 
complex hyperbolic space, with the negative curvature, there is a
bifurcation point at a critical value of the curvature which drastically 
changes the behavior of the system: whereas for small curvature 
(compared to the magnetic field) the dynamics
is periodic and bounded, at large curvature (compared to the
magnetic field) the dynamical trajectories become the geodesics of the complex
hyperbolic space which run to infinity.  
 In Sec. 4 we formulate a generalized model of Quantum Mechanics as a
Hamiltonian system with a magnetic field playing the role of the Hamiltonian.
We postulate that the {\it curvature of the Ka\"ahler manifold is determined by the
mass of the system} and, therefore, {\it large macroscopic systems exhibit 
an exponential rather than periodic behavior, which we interpret as the
collapse of the wave function}.

%\section{Curved Quantum Mechanics}
%=================================
%==================================================
%\section{Curved Quantum Mechanics}
%\setcounter{equation}{0} 
%=================================
%\section{Dynamics in Curved Quantum Mechanics}
%\setcounter{equation}{0} 
%===============================
%===================================================
\section{K\"ahler Manifolds}
%\section{Complex Geometry}
\setcounter{equation}{0} 
%\subsection{K\"ahler Manifolds}
%\subsection{Almost Complex Manifolds}
 
Let $(z^1,\dots,z^m, \bar z^1\dots, \bar z^m)$ be the 
holomorphic local coordinates
on an $2m$-dimensional almost complex manifold $M$.
We introduce Greek indices $\mu$ and $\bar\nu$ that run over
$1,\dots, m,$ and $\bar 1,\dots, \bar m$.
% as well as
%capital Latin indeces, $A,B,C$ that run over $1,\dots, m,$
%$ \bar 1,\dots, \bar m$.
We use the standard notation $\bar z^\mu=z^{\bar \mu}$
and adopt the notation
\be
\partial_\mu=\frac{\partial}{\partial z^\mu},
\qquad
\partial_{\bar\mu}=\frac{\partial}{\partial z^{\bar\mu}}.
\ee
An {almost complex structure} $J: T_pM\to T_pM$ 
is described by a real tensor of type $(1,1)$  
with the non-zero components
\be
J^\mu{}_\nu=i\delta^\mu{}_\nu,
\qquad
J^{\bar\mu}{}_{\bar \nu}=-i\delta^{\bar\mu}{}_{\bar\nu},
%\qquad
%I^{\mu}{}_{\bar \nu}=I^{\bar\mu}{}_{\nu}=0,
\ee

A Hermitian metric compatible with the almost complex structure is a tensor $g$
of type $(0,2)$ with the non-zero components $g_{\bar\mu\nu}$ and
$g_{\nu\bar\mu}=\overline{ g_{\mu \bar\nu}}$. The inverse metric is given by the
Hermitian matrix $g^{\nu\bar\mu}$. The Hermitian metric defines the interval
\be
ds^2=2g_{\mu\bar\nu}dz^\mu\; dz^{\bar\nu}
\ee
and the fundamental {\it K\"ahler form}
\be
\omega=ig_{\mu\bar\nu}dz^\mu\wedge dz^{\bar\nu}.
\ee
The corresponding volume form is
\bea
d\vol &=& \frac{1}{m!}\underbrace{\omega\wedge\cdots\wedge\omega}_m
\nn\\
&=&
i^m (\det g_{\mu\bar\nu})dz^1\wedge dz^{\bar 1}
\wedge\cdots\wedge dz^{m}\wedge dz^{\bar m}.
\eea

%\subsection{Fundamental Form}
%\setcounter{equation}{0} 
%===============================================================

Let $\nabla_X$ be the Hermitian connection which is compatible with the metric
and the almost complex structure.
Notice that, in general, the Hermitian connection is not torsion-free. 
A manifold is {\it Kähler} if it has 
three compatible structures: a complex structure, a Riemannian structure,
and a symplectic structure. In this case, the fundamental form is
symplectic, that is, non-degenerate and closed,
\be
d\omega=0,
\ee
the Hermitian
connection coincides with the Levi-Civita connection, that is, 
the torsion vanishes, 
and the almost complex structure is parallel with respect to the Levi-Civita
connection,
\be
\nabla_X J=0.
\ee
This also means that locally 
he metric satisfies the equations
\bea
\partial_\mu g_{\alpha\bar\beta} 
= \partial_\alpha g_{\mu\bar\beta},
\qquad
\partial_{\bar\mu} g_{\alpha\bar\beta} 
=\partial_{\bar\beta} g_{\alpha\bar\mu},
\eea
and, therefore, can be expressed as 
\be
g_{\mu\bar\nu}=\partial_\mu\partial_{\bar\nu} K,
\ee
where $K$ is a smooth real function called the K\"ahler potential.
So, the whole geometry is determined by a single function $K$.
%Of course, the K\"ahler potential is defined up to an arbitrary
%holomorphic function.
%\be
%K(z,\bar z)\mapsto K(z,\bar z)+f(z)+\overline{ f(z)}.
%\ee
%=========================
The only non-zero Christoffel coefficients are
\be
\Gamma^\nu{}_{\mu\alpha} 
= g^{\nu\bar\beta}\partial_\mu g_{\alpha\bar\beta},
\qquad
\Gamma^{\bar\nu}{}_{\bar\mu\bar\alpha} 
= g^{\beta\bar\nu}\partial_{\bar\mu} g_{\bar\alpha\beta},
\ee
and the equation of geodesics has the form
\bea
\frac{D \dot z^\mu}{dt}
=\ddot z^\mu
+\Gamma^{\mu}{}_{\alpha\beta}\dot z^\alpha\dot z^\beta
&=& 0.
\eea
The non-zero components of the Riemann tensor 
are
\bea
R^\alpha{}_{\beta\,\bar \mu\,\nu} &=& 
\partial_{\bar\mu}\left(g^{\alpha\bar \gamma}
\partial_\nu g_{\bar\gamma\beta}
\right).
\eea
The non-zero components of the Ricci tensor form a Hermitian
matrix, 
\bea
R_{\mu\bar\nu} &=& R^\alpha{}_{\mu\alpha\bar\nu}
+R^{\bar\alpha}{}_{\mu\bar\alpha\bar\nu}
\nn\\
&=& -\partial_{\mu}\partial_{\bar\nu}h,
%\\
%R_{\bar \nu \mu} &=& 
%\overline{R_{\nu \bar\mu}},
\eea
where
\be
h=\log\det g_{\mu\bar\nu}.
\ee
This enables one to define the {Ricci form}
\be
\cR=iR_{\mu\bar\nu}dz^\mu\wedge dz^{\bar\nu},
\ee 
which is a real closed two form,
\be
d\cR=0.
\ee
The manifold is {\it Einstein} if the Ricci tensor is proportional to the
metric
\be
R_{\mu\bar\nu}=\Lambda g_{\mu\bar\nu},
\ee
for some constant $\Lambda$,
and is called
{\it Calabi-Yau} if $\Lambda=0$.

%The scalar curvature is
%\bea
%R&=&
%-2g^{\mu\bar\nu}\partial_\mu\partial_{\bar\nu}h.
%\eea

%=====================================

In a neighborhood of every point, say, $z=0$, 
of a K\"ahler manifold there 
are normal holomorphic coordinates
such that the metric at the origin is Euclidean
\be
g_{\mu\bar\nu}(0)=\frac{1}{2}\delta_{\mu\nu},
\ee
and the connection at the origin is zero
\be
\Gamma^\mu{}_{\alpha\beta}(0)=0,
\ee
and, more generally, for any $k\ge 1$,
\be
\partial_{\alpha_1}\cdots\partial_{\alpha_k}
g_{\mu\bar\nu}(0) = 0.
\ee
The Taylor series for the metric near $z=0$ has the form
\be
g_{\mu\bar\nu}=\frac{1}{2}\delta_{\mu\nu}
+R_{\mu\bar\nu\bar\beta\alpha}(0)z^{\bar\beta}z^\alpha 
+O(z^3),
\ee
%and the potential near $z=0$ has the form
%\be
%K=\frac{1}{2}\rho 
%+ \frac{1}{4}V_{\bar\beta\alpha\bar\nu\mu}z^\alpha z^\mu  z^{\bar\beta}z^{\bar\nu}
%+O(z^5),
%\ee
%where
%$
%\rho=|z|^2=\la \bar z,z\ra
%$
%and 
%$V_{\bar\beta\alpha\bar\nu\mu}=R_{\bar\beta\alpha\bar\nu\mu}(0)$.

%======================================================

%========================================================= 

\subsection{Two-Dimensional K\"ahler Manifolds}
 
%=====================================================

These equations take a particularly simple form in the case of two dimensions,
that is, $m=1$. Every two-dimensional complex manifold is K\"ahler. In this case
the metric and the K\"ahler form are determined by one real positive function,
$g_{1\bar 1}=g$,
\bea
ds^2 &=& 2 g dz\;d\bar z,
\\
\omega &=& i g dz\wedge d\bar z.
\eea
The volume form is exactly equal to the the K\"ahler form
\be
d\vol=i g dz\wedge d\bar z.
\ee
In terms of the K\"ahler potential it is equal to
\be
g=\partial\bar\partial K,
\ee
where $\partial=\partial_z$,
The non-zero Christoffel coefficients are
\be
\Gamma^{1}{}_{11} =
\Gamma= \partial h,
\ee
where
\be
h=\log g,
\ee
and its conjugate.

The non-zero components of the Riemann tensor 
and the Ricci tensor are 
are
\be
R^1{}_{1\,\bar 1\,1}
=-R_{\bar 1 1}
=\partial\bar\partial h.
\ee
The scalar curvature is
\be
R=-2e^{-h}\partial\bar\partial h.
\ee
The manifold has constant curvature,
\be
R_{\bar 1 1}=\Lambda g, \qquad
R=2\Lambda,
\ee
with some constant $\Lambda$ if
\be
\partial\bar\partial \log g = -\Lambda g,
\ee
for some constant $\Lambda$.
It is not difficult to check that the function
\be
g=\frac{1}
{2\left(1+\frac{\Lambda}{4}z \bar z\right)^2}
\ee
is a solution of this equation.
For $\Lambda=0$ this is the standard metric on the complex plane $\CC^2$,
for $\Lambda>0$ this metric describes the sphere $S^2=\CC \PP^1$ of radius
$1/\sqrt{\Lambda}$;
for $\Lambda<0$ it is the metric of the Poincar\'e disc model of the 
hyperbolic space $H^2$ with pseudoradius $1/\sqrt{-\Lambda}$.
In the case $\Lambda>0$ the variables $z$ vary over the whole complex plane,
while for $\Lambda<0$ they vary over the disc
\be
|z|<\frac{2}{\sqrt{-\Lambda}}.
\ee

The Christoffel coefficients are
\bea
\Gamma &=& -\frac{\Lambda}{2}
\frac{\bar z}{\left(1+\frac{\Lambda}{4}z \bar z\right)},
\eea
and the equation of geodesics is
\be
\ddot z -\frac{\Lambda}{2}
\frac{\bar z\dot z^2}{\left(1+\frac{\Lambda}{4}z \bar z\right)}
=0,
\ee
with the integral of motion
\be
\frac{\dot z\overline{\dot z}}{\left(1+\frac{\Lambda}{4}z \bar z\right)^2}=C.
\ee

The geodesic with the initial conditions
\be
z(0)=0, \qquad \dot z(0)=u,
\ee
has the form
\be
z(t)=\frac{2}{\sqrt{\Lambda}}\tan\left(\frac{u}{2}\sqrt{\Lambda}t\right)
\qquad
\mbox{for } \Lambda>0,
\ee
and
\be
z(t)=\frac{2}{\sqrt{-\Lambda}}\tanh\left(\frac{u}{2}\sqrt{-\Lambda}t\right)
\qquad
\mbox{for } \Lambda\le 0.
\ee

In the case $\Lambda>0$ the manifold $\CC\PP^1=S^2$ is compact with the volume
\be
\vol(S^2)=\int \frac{idz\wedge d\bar z}{2\left(1+\frac{\Lambda}{4}z \bar 
z\right)^2}
=\frac{4\pi}{\Lambda}.
\ee
In the case $\Lambda\le 0$ the manifold $H^2$ is noncompact with infinite 
volume.

Now, let $\Phi(z)$ be an arbitrary nonconstant holomorphic function.
Then 
\be
\frac{\partial}{\partial z}\frac{\partial}{\partial\bar z}
=\partial\Phi\overline{\partial\Phi}
\frac{\partial}{\partial\Phi}\frac{\partial}{\partial\bar\Phi}.
\ee
Therefore, the general solution of this equation has the form
%\cite{polyanin2004}
\be
g=\frac{\partial\Phi \;\overline{\partial \Phi}}
{2\left(1+\frac{\Lambda}{4}\Phi\bar\Phi\right)^2}.
\ee
Therefore,
\be
ds^2=\frac{d\Phi \; d\bar \Phi}
{\left(1+\frac{\Lambda}{4}\Phi\bar\Phi\right)^2}.
\ee
Thus, for a general holomorphic function $\Phi(z)$ this is just a 
diffeomorphism.
That is why, the constant curvature manifolds are just $S^2$, $H^2$ and, 
of course, $\CC^2$.

%\subsection{Metric}
%\setcounter{equation}{0} 

%\subsection{Complex Projective Space $\CC\PP^m$}
%\setcounter{equation}{0} 
%===========================================

\subsection{K\"ahler Manifolds $\CC\PP^m$ and $\CC\HH^m$}

As an example we consider 
the complex projective space
$
\CC\PP^m = S^{2m+1}/U(1).
$
It is a K\"ahler manifold with the
K\"ahler potential
\be
K=\frac{a^2}{2}\log\left(1+\frac{\rho}{a^2}\right),
\ee
where $\rho=\la \bar z,z\ra$,
with a real parameter $a$;
the corresponding metric is the Fubini-Study metric 
on the  
\be
g_{\mu\bar\nu}=\frac{a^2}{2\left(a^2+\rho\right)^2}
\left[\left(a^2+\rho\right)\delta_{\mu\nu}
-z^{\bar\mu} z^\nu\right].
\label{424iga}
\ee
The the inverse metric is
\be
g^{\bar\nu\mu}
=\frac{2}{a^4}\left(a^2+\rho\right)\left(a^2\delta^{\nu\mu}
+z^{\bar\nu} z^\mu
\right).
\ee
%The determinant of the metric is
%\be
%e^h=\frac{a^{2(m+1)}}{2^m(a^2+\rho)^{m+1}}.
%\ee
%Note that the metric is invariant under the unitary transformation
%\be
%z^\mu \mapsto e^{i\alpha}z^\mu,
%\qquad
%\bar z^\mu \mapsto e^{-i\alpha}\bar z^\mu.
%\ee 
%In this case the function $f_2$ vanishes and 
The connection and the curvature tensors are
\bea
\Gamma^\beta{}_{\mu\nu} 
&=&
-\frac{1}{a^2+\rho}
\left({}z^{\bar\nu}\delta^\beta{}_{\mu}
+z^{\bar\mu} \delta^\beta{}_{\nu}
\right),
\\
%\bea
%R^\beta{}_{\mu\nu\bar\alpha}
%&=&
%\frac{2}{a^2}\left(\delta^\beta{}_\mu g_{\nu\bar\alpha}
%+\delta^\beta{}_\nu g_{\mu\bar\alpha}\right)
%\nn\\
%&=&
%\frac{1}{a^2+\rho}
%\left(\delta^\beta{}_{\mu}\delta_{\alpha\nu}
%+ \delta^\beta{}_{\nu}\delta_{\mu\alpha}
%\right)
%-\frac{1}{(a^2+\rho)^2}
%\left({}z^{\bar\nu}\delta^\beta{}_{\mu}
%+z^{\bar\mu} \delta^\beta{}_{\nu}
%\right)z^\alpha,
%\eea
%or
R_{\bar\beta\mu\nu\bar\alpha}
&=&
\frac{2}{a^2}\left(g_{\bar\beta\mu} g_{\nu\bar\alpha}
+g_{\bar\beta\nu} g_{\mu\bar\alpha}\right),
\\
R_{\mu\bar\nu} 
%=
%(m+1)\left( \frac{1}{\rho+2}\delta_{\mu\nu}
%-\frac{1}{(\rho+2)^2}z^\nu z^{\bar\mu}
%\right)
&=&\frac{2}{a^2}(m+1)g_{\mu\bar\nu}.
\eea
%{\bf CHECK the factor $2$} 
%This should be proportional to the metric!!!}
%The scalar curvature is
%\be
%R = \frac{4}{a^2} m (m+1).
%\ee

The geodesics are described by the equation
\be
\ddot z^\mu
-\frac{2}{a^2+\rho}{}\la \bar z, \dot z\ra\dot z^\mu=0
\ee
%Recall that here $\rho=|z|^2=z^\mu z^{\bar\mu}$.
%These equations can be solved exactly.
Since the manifold is homogeneous 
there is no distinguished point in,
the space looks the same at every point.
Therefore, without loss of generality we 
can choose the initial point to be the origin $z=0$, that is,
we impose the initial conditions
\be
z(0)=0,
\qquad
\dot z(0)=u.
\ee
There is the integral of these equations
\be
\frac{1}{2(a^2+\rho)^2}\left\{(a^2+\rho)|\dot z|^2
-|\la \bar z,\dot z\ra|^2
\right\}
=\frac{|u|^2}{2a^2}.
\ee
The solution is
\be
z^\mu(t)=r(t)\frac{u^\mu}{|u|},
\ee
where $r(t)$ satisfies the differential equation
\be
\ddot r-\frac{2}{a^2+r^2}r \dot r^2=0
\ee
with initial conditions
\be
r(0)=0, \qquad \dot r(0)=|u|.
\ee
The first integral has the form
\be
\dot r=|u|(a^2+r^2);
\ee
the solution of this equation is
\be
r(t)=a\tan\left(\frac{|u|}{a}t\right).
\ee

It is worth stressing that this equation is only local, valid
in a single coordinate chart. In fact, since the complex projective space
is compact, all geodesics are closed and have equal length.
The point $z=\infty$ corresponds to the antipodal point of the geodesic.
The arc length is
\be
ds=\frac{a^2}{a^2+r^2}dr.
\ee
Therefore, the length of the geodesic is twice the
distance to the antipodal point,
\be
L=2\int_0^\infty \frac{a^2}{a^2+r^2}dr
%=2a\tan^{-1}\left(\frac{r}{a}\right)\Big|_0^\infty
=\pi a.
\ee 

%==============================================

%\subsection{Complex Hyperbolic Space $\CC\HH^m$}

A similar potential
\be
K=-\frac{b^2}{2}\log\left(1-\frac{\rho}{b^2}\right),
\ee
corresponding to the imaginary $a=ib$,
leads to the metric
\be
g_{\mu\bar\nu}=\frac{b^2}{2\left(b^2-\rho\right)^2}
\left[\left(b^2-\rho\right)\delta_{\mu\nu}
+z^{\bar\mu} z^\nu\right],
\label{440iga}
\ee
with $\rho=|z|^2<b^2$,
which describes the complex hyperbolic space $\CC\HH^m$.
This manifold is noncompact, has negative curvature
\be
R_{\bar\beta\mu\nu\bar\alpha}
=
-\frac{2}{b^2}\left(g_{\bar\beta\mu} g_{\nu\bar\alpha}
+g_{\bar\beta\nu} g_{\mu\bar\alpha}\right),
\\
\ee
and infinite non-intersecting geodesics,
\be
z^\mu(t)=b\tanh\left(\frac{|u|}{b}t\right)\frac{u^\mu}{|u|}.
\ee

In the following we will make use of the product manifolds of two-dimensional
manifolds, that is, we assume that a general $2m$-dimensional manifold
$M$ is the product of $2$-dimensional submanifolds,
\be
M=M_1\times\cdots\times M_m,
\ee
with the metric
\be
ds^2=\sum_{\mu=1}^m 2 g_\mu(z^\mu,z^{\bar \mu})dz^\mu\; dz^{\bar \mu},
\ee
in particular,
\be
ds^2=\sum_{\mu=1}^m 
\frac{dz^\mu\; dz^{\bar \mu}}{\left(1+\frac{\Lambda_\mu}{4}z^\mu 
z^{\bar\mu}\right)^2},
\ee
where $\Lambda_\mu$ are real constants.
%(there is no summation over repeated indices!).

%===================================================================

\subsection{Infinite Dimensional K\"ahler Manifolds}

Our main idea is to generalize the standard Quantum Mechanics in the geometric
language. We replace the complex Hilbert space by an infinite-dimensional vector
bundle over an infinite-dimensional almost complex Hilbert manifold. In
particular, we consider the cotangent bundle $T^*\cM$ over a K\"ahler manifold
$\cM$ such that the cotangent space at every point is a Hilbert space. We will
deal with the infinite dimensional manifold rather formally as in finite
dimensions. We assume that the Riemannian metric defines the geodesic flow in
the usual way and that the manifold has a positive injectivity radius, that is,
the exponential map is well defined locally everywhere.

%\subsection{Infinite Dimensional Almost Complex Manifolds}
%\setcounter{equation}{0} 
%====================================
%  
%5/18/24
%
%====================================
%\item
%\end{enumerate}
%==================================================
%\section{Infinite Dimensional Almost Complex Manifolds}
%\setcounter{equation}{0} 
 
Let $M$ be a $n$-dimensional Riemannian manifold with a metric $g$
and $C^\infty(M)$ be the space of smooth complex valued functions on $M$.
We denote by 
\be
d\vol(x)=\sqrt{\det g_{\mu\nu}}\;dx^1\wedge\cdots\wedge dx^n
\ee
be the Riemannian volume element and introduce 
the standard pairing on $C^\infty(M)$,
\be
\la f,h\ra=\int\limits_M d\vol(x)\; f(x)h(x),
\ee
and the $L^2$ inner product
\be
(\psi,\varphi)=\la\bar\psi,\varphi\ra.
\ee

%We consider an infinite-dimensional almost complex Hilbert manifold $\cM$.
The local holomorphic coordinates on the Hilbert manifold
$\cM$ are smooth
complex functions $(z(x), \bar z(x))$ on 
$M$. 
Vector fields are variational operators acting on the 
functionals $\Phi(z,\bar z)$ on $\cM$. The tangent space $T_z\cM$
is spanned by the basis
$
\frac{\delta }{\delta z(x)}
$, $
\frac{\delta }{\delta \bar z(x)},
$
so that a vector field $\cX$ is a variational operator 
\bea
\cX(\Phi) &=&  
\left<X,\frac{\delta \Phi}{\delta z}\right>
+\left<\bar X,\frac{\delta\Phi}{\delta \bar z}\right>,
\eea
with components described by two smooth complex functions 
(which are functionals of $z$ and $\bar z$)
\be
\cX=\left(
\begin{array}{c}
X(x)\\
\bar X(x)
\end{array}
\right)
\in C^\infty(M)\oplus C^\infty(M).
\ee
%that is, $\cX\in C^\infty(M)\oplus C^\infty(M)$. 

One forms  are linear functionals on the vector fields.
The cotangent space $T_z^*\cM$ is spanned by the dual basis
$\delta z(x), \delta\bar z(x)$, so that a $1$-form $\psi$ is 
a functional 
%of the form
%\bea
%\psi &=& 
% \int\limits_M d\vol(x)\; \psi(x)\delta z(x)
%+ \int\limits_M d\vol(x)\; \bar\psi(x)\delta \bar z(x)
%\eea
with components
\be
\psi=\left(\psi(x),\bar\psi(x)\right) 
\in  C^\infty(M)\oplus C^\infty(M).
\ee
We slightly abuse notation by denoting the components of forms
in the same way as the $L^2$ product (this should not cause any
confusion).
We will omit the argument of the functions $X(x)$ and $\psi(x)$
where it does not cause any misunderstanding.
Then the value of a one-form $\psi$ on a vector $\cX$ is simply
given by the pairing
\bea
\psi(\cX) &=& \la\psi,X\ra+\la\bar\psi,\bar X\ra.
\eea

An almost complex structure is a map $J: T_z\cM\to T_z\cM$ defined by
\be
J\left(
\begin{array}{c}
X\\
\bar X
\end{array}
\right)
=\left(
\begin{array}{c}
iX\\
-i\bar X
\end{array}
\right).
\ee
Let $E: C^\infty(M)\to C^\infty(M)$ be an invertible operator
and $G{}$ be a self-adjoint operator defined by
\be
G{}=E^*E.
\ee
The inverse of the operator $G{}$ is
\be
G{}^{-1}=E^{-1} (E^{-1})^*
=\overline{D^*D},
\ee
where
\be
D=\overline{(E^{-1})^*}=(E^{-1})^T.
\ee
%========
This defines a Hermitian metric
$\cG$ on $\cM$ 
by
\bea
\cG(\cX,\cY) &=& \la \bar X,G{} Y\ra+\la\bar Y,G{} X\ra
\nn\\
&=&
(EX,EY)+(EY,EX).
\label{613iga}
\eea
which is compatible with the almost complex structure,
\be
\cG(J\cX,J\cY)=\cG(\cX,\cY).
\ee
Obviously,
\be
||\cX||^2=\cG(\cX,\cX)=2 (X,G{} X) = 2(EX,EX).
\ee
%======================
%so that
%\be
%\overline{G{}^{-1}}=D^*D.
%\ee
%======================
The inverse metric on the cotangent bundle
is defined by
\bea
\cG^{-1}(\psi,\varphi) &=& 
\la\psi,{G{}^{-1}}\bar\varphi\ra
+\la\varphi,{G{}^{-1}}\bar\psi\ra
\\
&=&(D\psi,D\varphi)+(D\varphi,D\psi),
\label{621iga}
\eea
so that
\be
||\psi||^2=\cG^{-1}(\psi,\psi)=2(D\psi,D\psi).
\ee
The fundamental K\"ahler $2$-form is defined by
\be
\omega(\cX,\cY)=\cG(J\cX,\cY)
=-i \la \bar X,G{} Y\ra+i\la\bar Y,G{} X\ra.
\ee

The Hermitian connection is the map 
\be
\cD_\cX: T\cM\to T\cM,
\ee
with $\cX$ being a vector,
that is compatible with the Hermitian metric, that is,
\be
\cG(\cD_\cX\cY,\cZ)+\cG(\cY,\cD_\cX\cZ)=\cX(\cG(\cY,\cZ)),
\ee
and with the almost complex structure
\be
\cD_\cX(J\cY)=J\cD_\cX\cY.
\ee
Let 
\be
\cX=\left(\begin{array}{c}
X\\
\bar X
\end{array}
\right),
\qquad
\cY=\left(\begin{array}{c}
Y\\
\bar Y
\end{array}
\right).
\ee
Then the vector
\be
\cZ=\cD_\cX\cY=\left(\begin{array}{c}
Z\\
\bar Z
\end{array}
\right)
\ee
is defined by
\bea
Z(x) &=& \left<X,\left(\frac{\delta }{\delta z}
+G{}^{-1}_x\frac{\delta G{}_x}{\delta z}\right)Y(x)\right>
\\
&=&
\left<X,\frac{\delta Y(x)}{\delta z}\right>
+\Gamma(x;X,Y),
\eea
where
\bea
\Gamma(x;X,Y) &=& G{}^{-1}_x\left<X,\frac{\delta G{}_x}{\delta z}\right>Y(x)
\\
&=&
\int\limits_M d\vol(y)\;X(y)G{}^{-1}_x\frac{\delta G{}_x}{\delta z(y)}Y(x).
\eea

%===========================
%\subsection{Infinite Dimenional K\"ahler Manifolds}

The manifold $\cM$ is K\"ahler if the torsion is zero,
that is,
\be
\Gamma(x;X,Y)=\Gamma(x;Y,X),
\ee
or
\be
\int\limits_M d\vol(y)\;X(y)\frac{\delta G{}_x}{\delta z(y)} Y(x)
=\int\limits_M d\vol(y)\;Y(y)\frac{\delta G{}_x}{\delta z(y)} X(x).
\ee
In this case the K\"ahler form $\omega$ is closed,
the Hermitian connection coincides
with the Levi-Civita connection.
This means that the operator $G{}$ is an integral operator
with the kernel $G{}(x,y)$,
\be
G{} X(x)=\int\limits_M d\vol(y)\; G{}(x,y)X(y),
\ee
satisfying the equation
\be
\frac{\delta G{}(u,x)}{\delta z(y)}
=\frac{\delta G{}(u,y)}{\delta z(x)}.
\ee
%===========================================
The kernel of the operator $G{}$ is 
determined by a potential functional $K=K(z,\bar z)$,
\be
G{}(x,y)=\frac{\delta^2 K}{\delta \bar z(x)\delta z(y)},
\ee
that is, the metric has the form
\be
\cG(\cX,\cY)= \int\limits_{M\times M} d\vol(x)\;d\vol(y)\; \left(
\bar X(x) G{}(x,y) Y(y)
+\bar Y(x) G{}(x,y) X(y)
\right).
\ee
 %====================================
The equation of geodesics is
\bea
\ddot z(x)
+\Gamma(x;{\dot z},\dot z)
=0.
\eea

%======================================================
%====================

We assume that the K\"ahler functional
near the origin has the form
\be
K=\frac{1}{2}\rho+O(z^4),
\ee
where
$
\rho=\la \bar z,z\ra,
$
so that the operator $G{}$ at the origin has the form
\be
G{}(0)=\frac{1}{2}I
\ee
and the connection is zero at the origin, that is,
\be
\frac{\delta G{}}{\delta z}(0)=0.
\ee

%==================================================================

%===========================
%\subsection{Complex Projective Spaces and Hyperbolic Spaces}
%\section{Examples}
%\setcounter{equation}{0} 
%==============================
%\subsection{Geodesics}
%\setcounter{equation}{0}  

%===============================

\subsection{Infinite Dimensional K\"ahler Manifolds 
$\CC\PP^\infty$ and $\CC\HH^\infty$}

We consider the infinite-dimensional complex projective space
$\CC\PP^\infty$ with the Fubini-Study metric
\be
\cG(\cX,\cY)=\frac{a^2}{2(a^2+\rho)}\left[\la\bar X,Y\ra+\la\bar Y,X\ra\right]
-\frac{a^2}{2(a^2+\rho)^2}\left[\la \bar X,z\ra\la\bar z,Y\ra
+\la\bar Y,z\ra\la\bar z,X\ra\right],
\ee
where
$
\rho=\la\bar z,z\ra,
$
with the K\"ahler potential
\be
K=\frac{a^2}{2}\log\left(1+\frac{\rho}{a^2}\right).
\ee
The operator $G{}$ has the form
\be
G{} X(x) = \frac{a^2}{2(a^2+\rho)}X(x)
-\frac{a^2}{2(a^2+\rho)^2}z(x)\la\bar z,X\ra.
\ee
The inverse metric is
\be
\cG^{-1}(\psi,\varphi)
=\frac{2}{a^4}(a^2+\rho)\left[
a^2\la\bar\psi,\varphi\ra+a^2\la\bar\varphi,\psi\ra
+\la \bar \psi,\bar z\ra \la z,\varphi\ra
+\la\bar\varphi,\bar z\ra\la z,\psi\ra
\right].
\ee
The inverse
 operator $G{}^{-1}$ has the form
\be
G{}^{-1} \varphi(x) = \frac{2}{a^4}(a^2+\rho)\left\{a^2\varphi(x)
+z(x)\la\bar z,\varphi\ra
\right\}.
\ee
The connection is
\bea
\Gamma(x;X,Y) 
&=&
-\frac{1}{a^2+\rho}
\left[X(x)\la\bar z,Y\ra
+Y(x)\la \bar z,X\ra
\right].
\eea
The equation of geodesics is
\bea
\ddot z(x)
-\frac{2}{a^2+\rho}
\la \bar z,\dot z\ra\dot z(x)=0.
\eea
The geodesic starting at the point $z(0)=0$ with the tangent
$u=\dot z(0)$ has the form
\be
z_t(x)=r(t)\frac{u(x)}{|u|},
\ee
where
\be
r(t)=a\tan\left(\frac{|u|}{a}t\right).
\ee
and $|u|^2=\la \bar u,u\ra$.

The infinite-dimensional complex hyperbolic space $\CC\HH^\infty$ is obtained
from $\CC\PP^\infty$ by changing the sign of the curvature, that is, by
replacing $a=ib$, so that $a^2=-b^2$,
\be
\cG(\cX,\cY)=\frac{b^2}{2(\rho-b^2)}\left[\la\bar X,Y\ra+\la\bar Y,X\ra\right]
+\frac{b^2}{2(\rho-b^2)^2}\left[\la \bar X,z\ra\la\bar z,Y\ra
+\la\bar Y,z\ra\la\bar z,X\ra\right],
\ee

%==========================================

%=======================================

\section{Hamiltonian Systems}
\setcounter{equation}{0} 

%=====================
\subsection{Einstein-K\"ahler Magnetic Field}
%=================================
%\setcounter{equation}{0} 

We also consider a closed two-form $\cF$
that we call a {\it magnetic field},
that is,
\be
d\cF=0.
\ee
Locally, it can be
represented by a potential one-form $\cA$,
\be
\cF=d\cA,
\ee
where
\be
\cA=\cA_++\cA_-
=\cA_\mu dz^\mu+\cA_{\bar\mu}dz^{\bar\mu}.
\ee
We assume that
\bea
\partial\cA_+ &=& \frac{1}{2}\left(\partial_\mu \cA_\nu
-\partial_\nu \cA_\mu\right)dz^\mu\wedge dz^\nu=0,
\\
\bar\partial\cA_-
&=& \frac{1}{2}\left(\partial_{\bar\mu} \cA_{\bar\nu}
-\partial_{\bar\nu} \cA_{\bar\mu}
\right)dz^{\bar\mu}\wedge dz^{\bar\nu}
=0,
\eea
therefore, the one-form $\cA$ is determined locally by
\be
\cA_\mu=-\frac{i}{2}\partial_\mu N,
\qquad
\cA_{\bar\mu}=\frac{i}{2}\partial_{\bar\mu} N,
\ee
with a real potential $N=N(z,\bar z)$,
so that
\be
\cA=-\frac{i}{2}(\partial-\bar\partial)N.
\ee
and the two-form $\cF$ is
\be
\cF
=iF_{\mu\bar\nu}dz^{\mu}\wedge dz^{\bar\nu},
\ee
where $F=(F_{\mu\bar\nu})$ is the Hermitian matrix 
\bea
F_{\mu\bar\nu} &=& \partial_\mu\partial_{\bar\nu}N
=N_{\mu\bar\nu}.
\eea

There are two closed forms on a K\"ahler manifold, the K\"ahler form,
$\omega$, and the Ricci form, $\cR$.
We assume that they are related by
\be
\cF=\varkappa\cR+\lambda \omega,
\ee
that is,
\be
F_{\mu\bar\nu}=\varkappa R_{\mu\bar\nu}+\lambda g_{\mu\bar\nu},
\ee
where $\lambda$ and $\varkappa$ are real parameters.
In this case
\be
N=\lambda K-\varkappa h.
\ee

If the manifold is Einstein, that is, 
$R_{\mu\bar\nu}=\Lambda g_{\mu\bar\nu}$, then
we have
\be
F_{\mu\bar\nu}=H g_{\mu\bar\nu},
\ee
where
\be
H=\varkappa \Lambda+\lambda,
\ee
and this form is parallel,
\be
\nabla\cF=0.
\ee
Such form will be called a {\it Einstein-K\"ahler magnetic field}.
In this case the tensor
\be
H_\mu{}^\nu=F_{\mu\bar\alpha}g^{\bar\alpha\nu}
\ee
is proportional to the identity,
\be
H_\mu{}^\nu
=H\delta_\mu{}^\nu.
\ee

Notice that in the case of a product of K\"ahler manifolds
\be
M=M_1\times \cdots\times M_n,
\label{589iga}
\ee
the corresponding K\"ahler magnetic fields $H_i$ on the submanifolds $M_i$
could be different. In this case we have
%\be
%g = \sum_{j=1}^n g_j,
%\ee
%where $g_j$ are the metrics on the submanifolds $\Sigma_j$, and
\be
\cF=\sum_{j=1}^n \left(\varkappa\cR_j+\lambda_j \omega_j\right),
\label{592iga}
\ee
where $\cR_j$ and $\omega_j$ are the Ricci tensor and the 
K\"ahler form on the submanifolds $\Sigma_j$.
In the case if all submanifolds $M_j$ are Einstein-K\"ahler, then
\be
\cF=\sum_{j=1}^n H_j \omega_j.
\ee
where
\be
H_j=\varkappa \Lambda_j+\lambda_j.
\ee

%==============================

As an example, for the product of two-dimensional manifolds of constant curvature,
$M_i$,
the tensor $F_{\mu\bar\nu}$ is diagonal,
\be
F_{\mu\bar\nu}=
\frac{1}{2\left(1+\frac{\Lambda_\mu}{4}z^\mu z^{\bar\mu}\right)^2}
H_\mu\delta_{\mu\nu},
\ee
(no summation over repeated indices here!)
and, therefore, the tensor $H$ is diagonal and constant,
\be
H_\mu{}^\nu=H_\mu\delta_\mu{}^\nu.
\ee

%STOP 
%9/14/2024
%xxxxxxxxxxxxxxxxxxxxxxxxxxxxxxxxx

\subsection{Hamiltonian Systems on K\"ahler Manifolds}

%============================================
 
The equations of geodesics on a K\"ahler manifold,
\be
\ddot z^\mu
+ \Gamma^{\mu}{}_{\alpha\beta} \dot z^{\alpha}\dot z^{\beta}=0,
\ee
can be written as a Hamiltonian system 
\bea
\frac{dz^\mu}{dt} &=& 
%\frac{\partial \cH}{\partial p_\mu}=
g^{\mu\bar\nu}p_{\bar\nu},
\\
%{\dot z^{\bar\mu}} &=& \frac{\partial \cH}{\partial p_{\bar\mu}}=g^{\bar\mu\nu} p_\nu,
%\\
\frac{dp_\nu}{dt} &=& 
%-\frac{\partial \cH}{\partial z^\nu}
%=-\frac{\partial g^{\alpha\bar\beta}}{\partial z^\nu}p_\alpha p_{\bar\beta},
%\nn\\
%&=& 
%=
\Gamma^{\alpha}{}_{\nu\gamma} g^{\gamma\bar\beta}p_{\alpha}  p_{\bar\beta},
\eea
with the Hamiltonian 
\be
\cH(z,\bar z,p,\bar p)=g^{\mu\bar\nu}(z,\bar z)p_\mu p_{\bar\nu},
\ee
(we use the notation $p_{\bar\mu}=\bar p_\mu$).
The directional derivative of a one-form along the curve $z(t)$
is described by
\be
\frac{D p_\nu}{dt}
=\nabla_{\dot z}p_\nu+\nabla_{\dot{\bar z}}p_\nu
=\frac{d p_\nu}{dt} 
- \Gamma^{\alpha}{}_{\nu\gamma} \dot z^{\gamma}p_{\alpha}.
\ee
Therefore, the momentum is parallel transported along the geodesics
%from the point $z(0)$ to the point $z(t)$,
\be
\frac{D p_\nu}{dt}=0
\ee
The Hamiltonian is an integral of the system, therefore, the
norm of the tangent vector is conserved,
\be
g_{\mu\bar\nu}(z,\bar z)\dot z^\mu{\dot z^{\bar\nu}}
=g^{\mu\bar\nu}(z,\bar z)p_\mu p_{\bar\nu}
=C.
\ee

%===========================================
%==============
%{\bf DO WE NEED THIS?}
%================
%===========================================

Notice, for a future reference that
the derivative of a two-form $X_{\mu\bar\nu}$ along a
curve $z(t)$ is determined by
\bea
\frac{D X_{\mu\bar\nu}}{dt}
&=&
\nabla_{\dot z} X_{\mu\bar\nu}
+\nabla_{\dot{\bar z}}X_{\mu\bar\nu}
\nn\\
&=&
\frac{d X_{\mu\bar\nu}}{dt}
-\Gamma^{\beta}{}_{\mu\alpha} \dot z^\alpha X_{\beta\bar\nu}
-\Gamma^{\bar\beta}{}_{\bar\nu\bar\alpha} \dot{z}^{\bar\alpha} X_{\mu\bar\beta}.
\eea

%========================================
%==========================================

We would like to emphasize the analogy with the dynamical
equations for a charged particle in a  magnetic field
on a Riemannian manifold. 
We consider a more general Hamiltonian system on a K\"ahler manifold
by taking into account a magnetic field (a closed 2-form $\cF$)
discussed above.
The Hamiltonian has the form
\bea
\cH(z,\bar z, p, \bar p) &=& \frac{1}{2}||\psi||^2
=g^{\mu\bar\nu}(z,\bar z)
\psi_{\mu}\psi_{\bar\nu},
\eea
where  
\be
\psi_\mu=p_\mu+q\cA_\mu,
\ee
and $q$ is a real parameter.
%==========================
The Hamiltonian system can be reduced to the system
\bea
\frac{dz^\mu}{dt} &=& {} g^{\mu \bar \nu}{}\psi_{\bar\nu},
\\
\frac{d\psi_\alpha}{dt} &=& 
%-\frac{e}{m}\cF_{\alpha\mu}g^{\mu\bar\nu}{}\psi_{\bar\nu}
-iqH_{\alpha}{}^\nu \psi_\nu
%-\frac{1}{m}\Gamma^{\mu}{}_{\alpha\beta}g^{\beta\bar\nu}
+{}\Gamma^{\mu}{}_{\alpha\beta}g^{\beta\bar\nu}
\psi_\mu {}\psi_{\bar\nu}.
%-\frac{\partial V}{\partial z^\alpha},
\eea
where $H_\mu{}^\nu=F_{\mu\bar\alpha}g^{\bar\alpha\nu}$.
It is easy to see that the system is invariant under the
gauge transformation
\be
\cA_\mu\mapsto \cA_\mu+\partial_\mu f,
\qquad
p_\mu \mapsto p_\mu - q\partial_\mu f.
\ee
This gives the second order equation
\bea
\nabla_{\dot z}\dot z^\mu
=\frac{D\dot z^\mu}{dt}
=\ddot z^\mu +\Gamma^\mu{}_{\alpha\beta}\dot z^\alpha \dot z^\beta
&=& iq H_{\beta}{}^\mu{\dot z^\beta}.
\eea
Of course, for $q=0$ this is just the equation of geodesics.
The equation for the momentum can be written in the 
parallel transport form too
\bea
\nabla_{\dot z}\psi_\alpha
=\frac{D\psi_\alpha}{dt}
=\frac{d\psi_\alpha}{dt} 
-\Gamma^{\nu}{}_{\alpha\beta}\dot z^\beta\psi_\nu {}
&=& 
-iqH_{\alpha}{}^\nu \psi_\nu.
\eea
Therefore, the parameter $q$ describes the deviation from the
parallel transport of the tangent vector and of the momentum.

%=======================================

This system has the integral of motion
\be
g_{\mu\bar\nu}\dot z^\mu\dot z^{\bar\nu}
= g^{\mu\bar\nu}
\psi_{\mu}\psi_{\bar\nu}
=C.
\ee
We impose the the initial conditions
\be
z(0)=0, \qquad
\xi=\psi(0).
\ee
Then the integral of motion is determined by
\be
C=2
|\xi|^2,
\ee
where  $|\xi|^2=\la\bar\xi,\xi\ra$.
%Therefore, for small $q$ the solution can be obtained by a perturbation theory 
%in the parameter $q$ with the zero order being the geodesics.
%For $q=0$ 
%this equation is exactly the equation of geodesics.
%Therefore, 

In the case when the manifold is the product of Einstein-K\"ahler manifolds
(\ref{589iga}) and the magnetic field is Einstein-K\"ahler (\ref{592iga}),
the Hamiltonian has the form
\be
\cH=\sum_{j=1}^n \cH_j(z_j,\bar z_j, p_j, \bar p_j)
\label{542iga}
\ee
and the Hamiltonian system decouples,
\bea
\frac{dz_j^\mu}{dt} &=& {} g_j^{\mu \bar \nu}{}\psi_{j,\bar\nu},
\label{543iga}
\\
\frac{d\psi_{j,\alpha}}{dt} &=& 
-iqH_{j} \psi_{j,\alpha}
-{}\frac{\partial g_j^{\mu \bar\nu}}{\partial z_j^\alpha}\psi_{j,\mu}
 {}\psi_{j,\bar\nu}.
\label{544iga}
\eea
with some constants $H_j$.
%This equation can be rewritten in the 
%parallel transport form
%\bea
%\nabla_{\dot z}\tilde\psi_{j,\alpha}
%=\frac{D\tilde\psi_{j,\alpha}}{dt}
%=\frac{d\tilde\psi_{j,\alpha}}{dt} 
%-\Gamma^{\nu}{}_{j,\alpha\beta}\dot z_j^\beta\tilde\psi_{j,\nu} {}
%&=& 0,
%\eea
%where
%\be
%\tilde\psi_{j,\alpha}=\exp(iqtH_j)\psi_{j,\alpha}.
%\ee 
The second-order equation reads
\bea
\ddot z_j^\mu +\Gamma_j^\mu{}_{\gamma\delta}\dot z_j^\gamma \dot z_j^\delta
&=& iq H_j{\dot z_j^\mu}.
\eea

%=======================================
%=======================================

% Fubini-Study metric
%============================================

\subsection{Hamiltonian Systems on Infinite Dimensional K\"ahler Manifolds}
%\setcounter{equation}{0} 

%\section{Hamiltonian Systems on Infinite Dimensional K\"ahler 
%Manifolds}
%\setcounter{equation}{0} 

We consider a Hamiltonian system on the cotangent bundle
$T^*\cM$ over a infinite-dimensional K\"ahler manifold $\cM$
with the local coordinates $(z,\psi)$, with $z\in \cM$ and $\psi\in T^*_z\cM$. 
We treat $\psi(t)$  as a co-tangent vector to a curve
$z(t)$ in a Hilbert manifold. We use the K\"ahler
metric (\ref{613iga})-(\ref{621iga}).
We consider a K\"ahler magnetic field  
\be
\cF=H\omega
\ee
with a potential $\cA$ given locally by
\be
\cA(x)=-\frac{i}{2}H\frac{\delta K}{\delta z(x)},
\qquad
\bar \cA(x)=\frac{i}{2}H\frac{\delta K}{\delta \bar z(x)},
\ee
where $H$ is a real constant.

%=============================
%=================================

We postulate the quantum dynamics to be the Hamiltonian system with the 
Hamiltonian
\bea
\cH(z,\bar z, \varphi, \bar \varphi) &=& 
{}\left<\psi,{G{}^{-1}}\bar\psi\right>,
%={}\left<\bar\psi,D^*D\psi\right>
%={}||D\psi||^2,
\eea
where 
\be
\psi=\varphi+iq\cA.
\ee
Then the Hamiltonian system takes the form
\bea
\frac{d z}{dt} &=& {} G{}^{-1}\bar\psi,
\\
\frac{d\psi}{dt} &=& 
-iq{} H\psi
-{}\left<\psi,\frac{\delta G{}^{-1}}{\delta z}\bar\psi\right>.
\eea
Recall that the operator $G{}$ is determined by the potential
functional
\bea
G{}(z;x,y) &=& \frac{\delta^2 K(z)}{\delta \bar z(x)\delta z(y)},
\eea
In more detail, these equations have the form
\bea
\frac{\partial z(x)}{\partial t} &=& {} 
\int\limits_M d\vol(y)\;G{}^{-1}(z;x,y)\bar\psi(y),
\\
\frac{\partial \psi(x)}{\partial t} &=& 
-iq{} H\psi(x)
-{}\int\limits_{M\times M} d\vol(y)\;d\vol(u)\; 
\psi(y)\frac{\delta G{}^{-1}(z;y,u)}{\delta z(x)}\bar\psi(u).
\eea
The corresponding 
second order equation has the form
\bea
\ddot z + \Gamma(\;\cdot\;;\dot z, \dot z) &=& 
iq{} H\dot z.
\eea
where
\bea
\Gamma(x;\dot z,\dot z) &=& G{}^{-1}_x
\left<\dot z,\frac{\delta G{}_x}{\delta z}\right>\dot z(x)
\\
&=&
\int\limits_M d\vol(y)\;\dot z(y)G{}^{-1}_x\frac{\delta G{}_x}{\delta z(y)}\dot z(x).
\eea
We impose the initial conditions
\be
z(0)=0,
\qquad
\psi(0)=\xi{}.
\ee
The initial velocity is determined by the initial momentum
\be
\dot z(0)=2{}\bar\xi{}.
\ee
This equation has an integral
\bea
\la\dot{\bar z},G{}\dot { z}\ra
=\la \psi,G{}^{-1}\bar\psi\ra
=|\xi|^2.
\eea

%=========================

%\section{Examples of Hamiltonian Systems}
%\setcounter{equation}{0} 

\subsection{Hamiltonian System in $\CC\PP^m$}
%\setcounter{equation}{0} 
%\setcounter{equation}{0} 

%==============================

The Hamiltonian system on complex projective space
$\CC\PP^m$ 
%and the complex hyperbolic space $\CC\HH^m$
with the Fubini-Study metric (\ref{424iga})
 and an Einstein-K\"ahler magnetic field
$H$ has the form
\bea
\frac{dz}{dt} &=& 2{} \left(1+\frac{\rho}{a^2}\right)\left({}\bar \psi
+\frac{1}{a^2}\la \bar z,\bar \psi\ra z
\right), 
\\
\frac{d\psi}{dt} &=& 
-iqH\psi
%\nonumber\\
%&&
-2\frac{{}}{a^2}\left(1+\frac{\rho}{a^2}\right)\la \bar z,\bar \psi\ra \psi
-2\frac{{}}{a^2}\left(|\psi|^2
+\frac{1}{a^2}|\la z,\psi\ra|^2 
\right)\bar z,
\eea
where $\rho=\la z,\bar z\ra$.
%Here $a^2>0$ for $\CC\PP^m$ and $a^2<0$ for $\CC\HH^m$. 
The second order equation has the form
\bea
\ddot z-\frac{2}{a^2+\rho}\la \bar z, \dot z\ra \dot z
=iqH\dot z.
\eea

We impose the initial conditions
\be
z(0)=0, 
\qquad \psi(0)=\xi,
\ee
which also means
\be
\dot z(0)=2{} \bar\xi.
\ee
The Hamiltonian integral of motion is
\bea
E &=& 2\frac{{}}{a^4}\left(a^2+|z|^2\right)
\left(a^2|\psi|^2+|\la z,\psi\ra|^2\right)
\nn\\
&=&\frac{a^2}{2{}(a^2+|z|^2)^2}
\left\{(a^2+\rho)|\dot z|^2-|\la \bar z,\dot z\ra|^2
\right\},
\eea
where
\be
E=2{}|\xi|^2.
\ee

For $q=0$ the solution of this equation 
is given by the geodesics
\bea
z(t) &=& a\tan\left(2{}\frac{|\xi|{}}{a}t\right)\frac{\bar \xi{}}{|\xi{}|},
\label{533gr}
\\[5pt]
\psi(t) &=& \cos^2\left(2{}\frac{|\xi|{}}{a}t\right) \xi{}.
\eea
%===============================
For $a\to \infty$, that is, zero curvature, we have $\dot z=0$ and, therefore,
$
z(t)=0.
$
Then the momentum is determined from the equation
\be
\frac{d\psi}{dt} = -iqH\psi,
\ee
which gives
\be
\psi(t)=U(t)\xi,
\ee
where
\be
U(t)=\exp\left(-iqtH\right).
\ee

In the general (when $qH\ne 0$ and a finite $a$)
case these equations are solved as follows.
The solution has the form
\bea
z(t) &=& a\zeta(t)\frac{\bar\xi}{|\xi|},
\\
\psi &=& \varphi(t)\xi.
\eea
Then the functions $\zeta(t)$ and $\varphi(t)$ 
satisfy the equations
\bea
\dot\zeta &=& 2\frac{{}|\xi|}{a} \left(1+\zeta\bar\zeta\right)^2\bar\varphi, 
\label{636iga}
\\
\dot\varphi &=& 
-iqH\varphi
-4\frac{{}|\xi|}{a}
\left(1+\zeta\bar\zeta\right)\bar\zeta\varphi\bar\varphi,
\eea
and
\be
\ddot \zeta -\frac{2}{1+\bar\zeta\zeta} \bar\zeta \dot \zeta^2 
=iqH\dot \zeta,
\label{638iga}
\ee
with the initial condition
\be
\zeta(0)=0,
\qquad
\varphi(0)=1,
\ee
and, therefore,
\be
\dot\zeta(0)=2\frac{{}|\xi|}{a}.
\ee
The Hamiltonian integral of motion is
\bea
2{}|\xi|^2 &=& 2{}|\xi|^2\left(1+\zeta\bar\zeta\right)^2
\varphi\bar\varphi
\nn\\
&=&\frac{a^2\dot\zeta\overline{\dot\zeta}}{2{}(1+\zeta\bar\zeta)^2}
\eea

Also, from the equation (\ref{638iga}) we see that
\be
\ddot\zeta(0)=2iqH\frac{{}|\xi|}{a}.
\ee
Therefore, the Taylor expansion of the function $\zeta(t)$ has the form
\be
\zeta(t)=2{} t+iqH\frac{{}|\xi|}{a} t^2+O(t^3).
\ee

Let
\be
\zeta=r e^{i\chi},
\ee
then
\bea
\dot \zeta &=&e^{i\chi}\left(
\dot r+ir\dot\chi\right),
\label{645iga}
\\
\ddot \zeta &=& e^{i\chi}\left(
\ddot r-r\dot\chi^2+2i\dot r\dot\chi+ir\ddot\chi\right).
\eea
This gives two equations
\bea
\ddot r
-\frac{2r}{1+r^2} \dot r^2
-\frac{1-r^2}{1+r^2}r\dot\chi^2
&=& -qH r\dot\chi,
\label{647iga}
\\
r\ddot\chi+2\frac{1-r^2}{1+r^2}\dot r\dot\chi
&=&qH\dot r.
\label{648iga}
\eea
In particular, we find
\be
\ddot r(0)=0.
\ee
To find the initial conditions for the functions $r(t)$ and $\chi(t)$
we use the Taylor series
\bea
r(t) &=& a_1 t +O(t^3),
\\
\chi(t) &=& b_1t+b_2t^2+O(t^3),
\eea
so that
\bea
\zeta(t) &=&
2\frac{{}|\xi|}{a} t+iqH\frac{{}|\xi|}{a} t^2 +O(t^3)
\nn\\
&=& a_1 t+i a_1b_1t^2+O(t^3),
\eea
which gives the initial conditions
\bea
&&r(0)=0
\qquad
\dot r(0)=2\frac{{}|\xi|}{a},
\\[5pt]
&&\chi(0)=0,
\qquad
\dot\chi(0)=\frac{qH}{2}.
%\frac{qHa^2}{2{}|\xi|}.
\eea

We introduce new variables 
\be
\gamma=\dot r,
\qquad
v=r\dot\chi
\ee
and use $r$ as an independent variable so that
\bea
\ddot r &=& \gamma\frac{d\gamma}{dr},
%=\frac{1}{2}\frac{d\gamma^2}{dr},
\\
\ddot\chi &=& \frac{\gamma}{r}\left(\frac{d\nu}{dr}-\frac{\nu}{r}\right). 
\eea
Then eqs. (\ref{647iga}), (\ref{648iga}) take the form
\bea
\frac{1}{2}\frac{d\gamma^2}{dr}
-\frac{2r}{1+r^2} \gamma^2
-\frac{1-r^2}{(1+r^2)r}\nu^2
&=& -qH \nu,
\label{647igam}
\\
\frac{d\nu}{dr}+\frac{1-3r^2}{r(1+r^2)}\nu
&=& qH.
\label{659iga}
\eea

%===========================

The solution of equation (\ref{659iga}) such that $v(0)=0$ is
\bea
v(r) &=& \frac{qH}{2}r(1+r^2), 
\eea
which gives
\be
\frac{d\chi}{dt}=\frac{qH}{2}(1+r^2).
\label{661iga}
\ee
The Hamiltonian integral of motion gives
\be
4\frac{{}|\xi|^2}{a^2}
%=2{}(1+r^2)^2\varphi\bar\varphi
=\frac{\dot r^2+r^2\dot\chi^2}{(1+r^2)^2}.
\ee
Therefore,
\bea
\dot r^2 &=&
4\frac{{}|\xi|^2}{a^2}(1+r^2)^2-r^2\dot\chi^2
\nn\\
&=& \frac{q^2H^2}{4}\left(k^2-r^2\right)(1+r^2)^2,
\eea
%
%======================================
%
% changed k to 1/k !!!!!!!!!!!!!!
%
%======================================
%%%%%
where
\be
k=\frac{4{}|\xi|}{qHa}.
\ee
Now, the function $r(t)$ is determined from the equation
\bea
\frac{dr}{dt} &=& \frac{qH}{2}(1+r^2)\sqrt{k^2-r^2}
\label{665iga}
\eea
with the initial condition $r(0)=0$.
The solution of this equation is given implicitly by
\bea
\omega t &=& \tan^{-1}\left(
\sqrt{1+k^2}\frac{r}{\sqrt{k^2-r^2}}
\right),
\eea
where
\bea
\omega 
&=&\sqrt{4\frac{{}|\xi{}|^2}{a^2}+\frac{q^2H^2}{4}}
\nn\\
&=& \frac{qH}{2}\sqrt{1+k^2}.
\eea
The explicit solution has a very simple form
\be
r(t)=\frac{k\sin\left(\omega t\right)}
{\sqrt{1+k^2\cos^2\left(\omega t\right)}}.
\label{668iga}
\ee
The function $\chi(t)$ is determined from the equation
(\ref{661iga}), which, by using (\ref{668iga}), gives
\bea
\frac{d\chi}{dt} &=& \frac{qH}{2}(1+r^2)
\nn\\
&=&
\frac{qH}{2}(1+k^2)\frac{1}{1+k^2\cos^2(\omega t)}.
\label{669iga}
\eea
The solution with the initial condition $\chi(0)=0$ has the form
\bea
\chi(t) &=& 
\tan^{-1}\left(
\frac{\tan\left(\omega t\right)}{\sqrt{1+k^2}}
\right).
\eea
Further, we compute
%\bea
%\cos\chi &=& \frac{\sqrt{1+k^2}\cos(\omega t)}{\sqrt{k^2+\cos^2(\omega t)}},
%\\
%\sin\chi &=& \frac{k\sin(\omega t)}{\sqrt{k^2+\cos^2(\omega t)}},
%\eea
%and, therefore,
\be
e^{i\chi}=
 \frac{\sqrt{1+k^2}\cos(\omega t)
+i\sin(\omega t)}{\sqrt{1+k^2\cos^2(\omega t)}},
\ee
and, therefore, by using (\ref{668iga})
we obtain
\bea
\zeta(t) &=& 
%re^{i\chi}
%\nn\\
%&=&
k\frac{\sin(\omega t)}{1+k^2\cos^2(\omega t)}
\left\{
\sqrt{1+k^2}\cos(\omega t)
+i\sin (\omega t)
\right\}.
\eea
Next, by using (\ref{645iga}), (\ref{665iga}) and (\ref{669iga}) 
we compute the derivative
\be
\dot\zeta=\frac{qH}{2}
\left\{\sqrt{k^2-r^2}
+ir\right\}\frac{(1+r^2)}{r}\zeta,
\ee
and, by using (\ref{636iga}) and (\ref{668iga}), the function $\varphi$,
\bea
\varphi(t) &=&
\left\{
\cos(\omega t)-i\frac{1}{\sqrt{1+k^2}}\sin(\omega t)\right\}^2.
\eea

Thus, the dynamics of the Hamiltonian system is given by
\bea
z(t) &=& a k\sqrt{1+k^2}
\frac{\sin(\omega t)}{\left[1+k^2\cos^2(\omega t)\right]}
\left\{
\cos(\omega t)
+i\frac{1}{\sqrt{1+k^2}}\sin (\omega t)
\right\}\frac{\bar \xi}{|\xi|},
\nn\\
&=&
 a k
\frac{
i\left[1-\cos(2\omega t)\right]
+\sqrt{1+k^2}\sin(2\omega t)
}{2+k^2+k^2\cos(2\omega t)}
\frac{\bar \xi}{|\xi|},
\\
\psi(t) &=&
\left\{
\cos(\omega t)-i\frac{1}{\sqrt{1+k^2}}\sin(\omega t)\right\}^2
\xi{}
\\
&=&
\frac{1}{2}\left\{
\frac{k^2}{1+k^2}+\frac{2+k^2}{1+k^2}\cos(2\omega t)
-2i\frac{1}{\sqrt{1+k^2}}\sin(2\omega t)\right\}
\xi
\nn
\eea
It is easy to see that
\bea
|z(t)|^2 &=& a^2k^2
\frac{\sin^2(\omega t)}{[1+k^2\cos^2(\omega t)]}
%\le 16\frac{{}|\xi|^2}{q^2H^2},
\\
|\psi(t)|^2 &=& \frac{|\xi|^2}{(1+k^2)^2}\left(
1+k^2\cos^2(\omega t)\right)^2
%\le \frac{|\xi|^2}{(1+k^2)^2},
\\
|\la z,\psi\ra|^2 &=&
a^2|\xi|^2\frac{k^2}{(1+k^2)^2}
\sin^2(\omega t)\left(
1+k^2\cos^2(\omega t)\right)
\eea
Further,
\bea
a^2+|z|^2 &=& 
\frac{a^2(1+k^2)}{1+k^2\cos^2(\omega t)}
\\
a^2|\psi|^2+|\la z,\psi\ra|^2 &=&
% \frac{a^2|\xi|^2}{(1+k^2)^2}\left(
%1+k^2\cos^2(\omega t)\right)^2
%\\
%&&+a^2|\xi|^2\frac{k^2}{(1+k^2)^2}
%\sin^2(\omega t)\left(
%1+k^2\cos^2(\omega t)\right)
%\\
%&=&
\frac{a^2|\xi|^2}{(1+k^2)}
\left(1+k^2\cos^2(\omega t)\right)
\eea
This gives the norm
\be
||\psi||^2=\frac{4}{a^4}\left(a^2+|z|^2\right)\left(a^2|\psi|^2+|\la 
z,\psi\ra|^2\right)
=4|\xi|^2,
\ee
and the Hamiltonian integral
\bea
E=\frac{{}}{2}||\psi||^2
=2{}|\xi|^2.
\eea

%==========================

It is easy to see that for any non-zero $qH\ne 0$ 
the trajectories are smooth and exist for any time. They are periodic
with the period
\be
T=\frac{\pi}{\omega};
%=\frac{\pi}{\sqrt{{}|\xi{}|^2+\frac{q^2H^2}{16}}}.
\ee
and the length
\be
L
=\frac{a\pi k}{\sqrt{1+k^2}}.
\ee
However, for $q=0$ (or $k\to \infty$) the trajectories blow up at a finite time 
\be
t_*=\frac{a\pi}{4{} |\xi{}|}.
\ee

%\marginpar{\bf COMPUTE MOMENTUM}
%

There are two limiting cases.

{\bf Case I:}  $k\to 0$.
In this case the dynamical trajectories are
\bea
z(t) &=& 
\frac{4{}}{qH}\left\{
1-\frac{k^2}{2}\cos(\omega t)\exp(-i\omega t)
+O\left(k^4\right)
\right\}\sin(\omega t)\exp(i\omega t)\bar\xi,
\nn\\
\\
\psi(t) &=& \left\{1+ik^2\sin(\omega t)\exp(i\omega t)
+O\left(k^4\right)\right\}\exp(-2i\omega t)\xi.
\eea
Of course, in the limit $k\to 0$ we recover
\bea
z(t) &=& 
\frac{4{}}{qH}\sin\left(\frac{qH}{2}t\right)
\exp\left(i\frac{qH}{2} t\right)\bar \xi{},
\\
\psi(t) &=&  \exp\left(-iqH t\right)\xi.
\eea
Notice that this asymptotic solution does not depend on the parameter
$a$.

{\bf Case II:} 
$k\to \infty$. 
In this case the dynamical trajectories are 
\bea
z(t) &=& a
\left\{
1+\frac{i}{k}\tan(\omega t) + O\left(\frac{1}{k^2}\right)
\right\}
\tan\left(\omega t\right)\frac{\bar \xi{}}{|\xi{}|},
\\
\psi(t) &=& 
\left\{
1-\frac{2i}{k}\tan(\omega t)+O\left(\frac{1}{k^3}\right)
\right\}
\cos^2\left(\omega t\right)\xi{}.
\eea
In the limit $k\to \infty$ they are
the geodesics
of the Fubini-Study metric on the complex projective space $\CC\PP^m$,
\bea
z(t) &=& a\tan\left(2\frac{{}|\xi{}|}{a} t\right)\frac{\bar \xi{}}{|\xi{}|},
\\
\psi(t) &=& 
\cos^2\left(2\frac{{}|\xi{}|}{a} t\right)\xi{}.
\eea

%xxxxxxxxxxxxxxxxxxxxxxxxxxxxxxxxxxx
%STOP
%9/16/24
%xxxxxxxxxxxxxxxxxxxxxxxxxxxxxxxxxxxxx

%========================================================

\subsection{Hamiltonian System in $\CC\HH^m$}

The Hamiltonian system on the complex hyperbolic space $\CC\HH^m$
with the Fubini-Study type metric
(\ref{440iga}) and an Einstein-K\"ahler magnetic field
$H$ is obtained from the system for the complex projective space
$\CC\PP^m$ by formally changing the sign of $a^2=-b^2$,
\bea
\frac{dz}{dt} &=& 2{} \left(1-\frac{\rho}{b^2}\right)\left({}\bar \psi
-\frac{1}{b^2}\la \bar z,\bar \psi\ra z
\right), 
\label{5122iga}
\\
\frac{d\psi}{dt} &=& 
-iqH\psi
+2\frac{{}}{b^2}\left(1-\frac{\rho}{b^2}\right)\la \bar z,\bar \psi\ra \psi
+2\frac{{}}{b^2}\left(|\psi|^2
-\frac{1}{b^2}|\la z,\psi\ra|^2 
\right)\bar z,
\label{5123iga}
\eea
where $\rho=\la z,\bar z\ra$.
The second order equation has the form
\bea
\ddot z+\frac{2}{b^2-\rho}\la \bar z, \dot z\ra \dot z
=iqH\dot z.
\eea

We impose the initial conditions
\be
z(0)=0, 
\qquad \psi(0)=\xi,
\ee
which also means
\be
\dot z(0)=2{} \bar\xi.
\ee
The Hamiltonian integral of motion is
\bea
E &=& 2\frac{{}}{b^4}\left(b^2-|z|^2\right)
\left(b^2|\psi|^2-|\la z,\psi\ra|^2\right)
\nn\\
&=&\frac{b^2}{2{}(b^2-|z|^2)^2}
\left\{(b^2-\rho)|\dot z|^2+|\la \bar z,\dot z\ra|^2
\right\}
\eea
where
\be
E=2{}|\xi|^2.
\ee

These equation can be solved in a similar way.
We define
\bea
k &=&\frac{4{}|\xi|}{qHb},
\\
\omega &=&\sqrt{\frac{q^2H^2}{4}-4\frac{{}|\xi|^2}{b^2}}
\nn\\
&=&
\frac{qH}{2}\sqrt{1-k^2}.
\eea

In the case of the negative curvature the parameter $k$ plays a much
more crucial role. The behavior of the system is very different
for $k<1$ and for $k>1$ with a singularity at $k=1$.
We consider first the case $k<1$.
Then the solution has qualitatively the same form
\bea
z(t) &=& bk\sqrt{1-k^2}
\frac{\sin(\omega t)}{[1-k^2\cos^2(\omega t)]}
\left\{
\cos(\omega t)
+i\frac{1}{\sqrt{1-k^2}}\sin (\omega t)
\right\}\frac{\bar \xi}{|\xi|},
\nn\\
&=&
 b k
\frac{
i\left[1-\cos(2\omega t)\right]
+\sqrt{1-k^2}\sin(2\omega t)
}{2-k^2-k^2\cos(2\omega t)}
\frac{\bar \xi}{|\xi|},
\label{5131iga}
\\
\psi(t) &=&
\left\{
\cos(\omega t)-i\frac{1}{\sqrt{1-k^2}}\sin(\omega t)\right\}^2
\xi{}
\nn\\
&=&
\frac{1}{2}\left\{
-\frac{k^2}{1-k^2}+\frac{2-k^2}{1-k^2}\cos(2\omega t)
-2i\frac{1}{\sqrt{1-k^2}}\sin(2\omega t)\right\}
\xi
\label{5132iga}
\eea
These trajectories are still periodic with the period
\be
T=\frac{\pi}{\omega}.
\ee
The norms of the solution are
\bea
|z(t)|^2 &=& b^2k^2
\frac{\sin^2(\omega t)}{1-k^2\cos^2(\omega t)}
%\le \frac{16{}|\xi|^2}{q^2H^2}
%\frac{1}{1-k^2},
\\
|\psi(t)|^2 &=& \frac{|\xi|^2}{(1-k^2)^2}\left(
1-k^2\cos^2(\omega t)\right)^2
%\le \frac{|\xi|^2}{1-k^2}.
\eea 
and
\be
||\psi||^2=4|\xi|^2.
\ee
As $k\to 0$ we have 
\bea
z(t) &=& 
\frac{4{}}{qH}\left\{
1+\frac{k^2}{2}\cos(\omega t)\exp(-i\omega t)
+O\left(k^4\right)
\right\}\sin(\omega t)\exp(i\omega t)\bar\xi,
\nn\\
\\
\psi(t) &=& \left\{1-ik^2\sin(\omega t)\exp(i\omega t)
+O\left(k^4\right)\right\}\exp(-2i\omega t)\xi,
\label{5138iga}
\eea
with the same limit
\bea
z(t) &=& 
\frac{4{}}{qH}\sin\left(\frac{qH}{2}t\right)
\exp\left(i\frac{qH}{2} t\right)\bar \xi{},
\\
\psi(t) &=&  \exp\left(-iqHt\right)\xi.
\eea
We call this behavior the {\it quantum regime}.

Thus, the impact of the curvature is in the frequency $\omega$. Positive
curvature increases the frequency, whereas the negative curvature decreases it.
As the parameter $k$ increases from $0$ to $1$, the frequency decreases until it
becomes zero at $k=1$. In the case $k>1$ the frequency becomes imaginary
\be
\omega=i\mu,
\ee
with
\be
\mu=\frac{qH}{2}\sqrt{k^2-1},
\ee
and the solution is not periodic
any longer.

For $k>1$ the solution has a very different exponential form
\bea
z(t) &=& bk\sqrt{k^2-1}
\frac{\sinh(\mu t)}{[k^2\cosh^2(\mu t)-1]}
\left\{
\cosh(\mu t)
+i\frac{1}{\sqrt{k^2-1}}\sinh (\mu t)
\right\}\frac{\bar \xi}{|\xi|},
\nn\\
&=&
b k
\frac{
i\left[\cosh(2\mu t)-1\right]
+\sqrt{k^2-1}\sinh(2\mu t)
}{k^2\cosh(2\mu t)+k^2-2}
\frac{\bar \xi}{|\xi|},
\label{5143iga}\\
\psi(t) &=&
\left\{
\cosh(\mu t)-i\frac{1}{\sqrt{k^2-1}}\sinh(\mu t)\right\}^2
\xi{}.
\nn\\
&=&
\frac{1}{2}\left\{
\frac{k^2}{k^2-1}+\frac{k^2-2}{k^2-1}\cosh(2\mu t)
-2i\frac{1}{\sqrt{k^2-1}}\sinh(2\mu t)\right\}
\xi
\label{5144iga}
\eea
This solution exists for any $t$. As $t\to \infty$,
$z(t)$ approaches a fixed point,
\be
z(t)\to z_\infty= 
b\frac{\sqrt{k^2-1}+i}{k}\frac{\bar \xi}{|\xi|},
\ee
with
\be
|z_\infty|=b,
\ee
which corresponds to the point at infinity of the 
hyperbolic space.
The function $\psi(t)$ grows exponentially 
without bound as $t\to \infty$,
\be
\psi(t) \sim
\frac{1}{4(k^2-1)}\left\{
k^2-2
-2i\sqrt{k^2-1}\right\}
\exp\left(2\mu t\right)
\xi,
\ee 
with 
\be
|\psi(t)|=\frac{k^2}{4(k^2-1)}
\exp\left(2\mu t\right)
|\xi|.
\ee

As $k\to \infty$ the dynamical trajectories are 
\bea
z(t) &=& b\left\{
1+\frac{i}{k}\tanh(\mu t) + O\left(\frac{1}{k^2}\right)
\right\}
\tanh\left(\mu t\right)\frac{\bar \xi}{|\xi|},
\\
\psi(t) &=& 
\left\{
1-\frac{2i}{k}\tanh(\mu t)+O\left(\frac{1}{k^3}\right)
\right\}
\cosh^2\left(\mu t\right)\xi{},
\eea
so that in the limit $k\to \infty$ they are
the geodesics
on the complex hyperbolic space $\CC\HH^m$,
\bea
z(t) &=& b\tanh\left(2\frac{{}|\xi{}|}{b} t\right)\frac{\bar 
\xi{}}{|\xi{}|},
\\
\psi(t) &=& 
\cosh^2\left(2\frac{{}|\xi{}|}{b} t\right)\xi{}.
\eea

We call this behavior the {\it classical regime} (or the geometric one).
Thus, the curvature ${}$ provides an interpolation
between the quantum and the classical regime. 
The quantum world becomes classical at the bifurcation point
when periodic solutions become not periodic.

%It is interesting to study the critical regime $k=1$.
%===================================================================
%==========================
%STOP 11/18/24
%==========================

%=======================================================
%\section{A Model of Curved Quantum Mechanics}
%\setcounter{equation}{0} 
%==========================================
%\section{Hamiltonian System in $\CC\PP^\infty$ and $\CC\HH^\infty$}
%\setcounter{equation}{0} 
%=========================

%====================================================
%=======================================================
%\section{A Model of Curved Quantum Mechanics}
%\setcounter{equation}{0} 
%==========================================
%\subsection{Hamiltonian System in $\CC\PP^\infty$}
%=========================

For the complex projective manifold $\CC\PP^\infty$ 
with the Fubini-Study metric the
Hamiltonian system is
\bea
\frac{dz}{dt} &=& 2{}\left(1+\frac{\rho}{a^2}\right)
\left(
\bar\psi+\frac{1}{a^2}z\la\bar z,\bar\psi\ra
\right),
\\
\frac{d\psi}{dt} &=& 
-iq{}H\psi
%\\
%&&
-2\frac{1}{a^2}\left(1+\frac{\rho}{a^2}\right)\la \bar z,\bar \psi\ra \psi
-2\frac{1}{a^2}\left[\la\bar\psi,\psi\ra
+\frac{1}{a^2}\la\bar z,\bar\psi\ra \la z,\psi\ra
\right]\bar z,
\nn\\
\eea

This system with the initial conditions
\be
z(0)=0, \qquad
\psi(0)=\xi,
\ee
is solved exactly in the same way as in finite dimensions;
the solution has the form
\bea
z(t) &=& 
 a k
\frac{
i\left[1-\cos(2\omega t)\right]
+\sqrt{1+k^2}\sin(2\omega t)
}{2+k^2+k^2\cos(2\omega t)}
\frac{\bar \xi}{|\xi|},
\\
\psi(t) &=&
\frac{1}{2}\left\{
\frac{k^2}{1+k^2}+\frac{2+k^2}{1+k^2}\cos(2\omega t)
-2i\frac{1}{\sqrt{1+k^2}}\sin(2\omega t)\right\}
\xi
\nn
\eea
where
\bea
k&=&\frac{4{}|\xi|}{qHa},
\\
\omega 
%&=&\sqrt{4\frac{|\xi{}|^2}{a^2}+\frac{q^2H^2}{4}}
%\nn\\
&=& \frac{qH}{2}\sqrt{1+k^2}.
\eea

For the complex hyperbolic space there are two different regimes.
Let
\bea
k&=&\frac{4{}|\xi|}{qHb},
\\
\omega 
%&=&\sqrt{-4\frac{|\xi{}|^2}{b^2}+\frac{q^2H^2}{4}}
%\nn\\
&=& \frac{qH}{2}\sqrt{1-k^2}.
\eea
For $k<1$ the solution has the form
\bea
z(t) &=& 
 b k
\frac{
i\left[1-\cos(2\omega t)\right]
+\sqrt{1-k^2}\sin(2\omega t)
}{2-k^2-k^2\cos(2\omega t)}
\frac{\bar \xi}{|\xi|},
\label{5131iga}
\\
\psi(t) &=&
\frac{1}{2}\left\{
-\frac{k^2}{1-k^2}+\frac{2-k^2}{1-k^2}\cos(2\omega t)
-2i\frac{1}{\sqrt{1-k^2}}\sin(2\omega t)\right\}
\xi
\label{5132igax}
\eea
These trajectories are periodic with the period
\be
T=\frac{\pi}{\omega}.
\ee
For small $k$ we have 
\bea
z(t) &=& 
\frac{4{}}{qH}\left\{
1+\frac{k^2}{2}\cos(\omega t)\exp(-i\omega t)
+O\left(k^4\right)
\right\}\sin(\omega t)\exp(i\omega t)\bar\xi,
\nn\\
\\
\psi(t) &=& \left\{1-ik^2\sin(\omega t)\exp(i\omega t)
+O\left(k^4\right)\right\}\exp(-2i\omega t)\xi,
\eea
with the limit as $k\to 0$
\bea
z(t) &=& 
\frac{4{}}{qH}\sin\left(\frac{qH}{2}t\right)
\exp\left(i\frac{qH}{2} t\right)\bar \xi{},
\\
\psi(t) &=&  \exp\left(-iqHt\right)\xi,
\eea
We call this behavior the {\it quantum regime}.

In the case $k>1$ the frequency becomes imaginary
\be
\omega=i\mu,
\ee
with
\be
\mu=\frac{qH}{2}\sqrt{k^2-1},
\ee
and the solution has the form
\bea
z(t) &=& 
b k
\frac{
i\left[\cosh(2\mu t)-1\right]
+\sqrt{k^2-1}\sinh(2\mu t)
}{k^2\cosh(2\mu t)+k^2-2}
\frac{\bar \xi}{|\xi|},
\label{5143iga}\\
\psi(t) &=&
\frac{1}{2}\left\{
\frac{k^2}{k^2-1}+\frac{k^2-2}{k^2-1}\cosh(2\mu t)
-2i\frac{1}{\sqrt{k^2-1}}\sinh(2\mu t)\right\}
\xi
\label{5144igax}
\eea
So, it is not periodic
any longer. 
For large $k$ it takes the form
\bea
z(t) &=& b
\left\{
1+\frac{i}{k}\tanh(\mu t) + O\left(\frac{1}{k^2}\right)
\right\}
\tanh\left(\mu t\right)\frac{\bar \xi}{|\xi|},
\\
\psi(t) &=& 
\left\{
1-\frac{2i}{k}\tanh(\mu t)+O\left(\frac{1}{k^3}\right)
\right\}
\cosh^2\left(\mu t\right)\xi{},
\eea
which become the geodesics
on the complex hyperbolic space $\CC\HH^\infty$ as $k\to \infty$,
\bea
z(t) &=& b\tanh\left(2\frac{{}|\xi{}|}{b} t\right)\frac{\bar \xi{}}{|\xi{}|},
\\
\psi(t) &=& 
\cosh^2\left(2\frac{{}|\xi{}|}{b} t\right)\xi{}.
\eea
This solution exists for any $t$. 
The function $\psi(t)$ grows exponentially 
without bound as $t\to \infty$,
\be
|\psi(t)| \sim \frac{k^2}{4(k^2-1)}
\exp(2\mu t)|\xi|.
\ee 
We call this behavior the {\it classical regime} (or the geometric one).
Thus, the curvature ${}$ provides an interpolation
between the quantum and the classical regime. 
The quantum world becomes classical
when the periodic solutions become not periodic.

%===========================
%============================================

\section{Curved Quantum Mechanics}
\setcounter{equation}{0} 

%========================
%xxxxxxxxxxxxxxxxxxxxxxxxxxxxxxxx
%=========================

We start with the usual Hamiltonian operator of Quantum Mechanics.
Let 
\be
H: \cV\to \cV
\ee
be a self-adjoint operator on the 
Hilbert space $\cV=L^2(M)$ bounded from below. Without loss of generality, 
we can always assume it to be positive.
We suppose that it has a discrete 
real spectrum $(H_j)_{j=1}^\infty$ bounded from below with finite multiplicities
$d_j$. Let $V_j$ be the corresponding complex finite-dimensional eigenspaces
of dimension $d_j$ and $P_j$ be the corresponding projections
so that
\be
H=\sum_{j=1}^\infty H_j P_j.
\ee
We identify these finite-dimensional vector spaces with
the cotangent spaces 
\be
V_j=T_z^* M_j
\ee
to 
finite-dimensional K\"ahler manifolds $M_j$, in particular, the complex
hyperbolic spaces $M_j=\CC\HH^{d_j}$ of dimension $d_j$
with the Fubini-Study type metric (\ref{440iga}) with
the negative curvature 
\be
\Lambda_j=-\frac{1}{4b_j^2}
%=-\frac{G^2m^6}{4\hbar^4}
\ee
determined by the radius $b_j$.
%\be
%b_j=\frac{\hbar^2 }{Gm^3}.
%\ee 
For simplicity, we restrict ourselves to the nondegenerate case
when the eigenvalues of the operator $H$ are simple,
the submanifolds $M_j$ are two dimensional, 
\be
M_j=\CC\HH^1=H^2,
\ee
that is, are just the hyperbolic planes.
This defines the infinite-dimensional state manifold
\be
\cM=\varprod_{j=1}^\infty M_j
\ee
with the cotangent space
\be
\cV=T_z^*\cM=\bigoplus_{j=1}^\infty T_z^*M_j
\ee

%=========================================

Let $\varphi_j=\varphi_j(x)$ be the orthonormal basis of eigenvectors of the operator $H$
for the Hilbert space $\cV$ so that
\be
P_j=|\varphi_j\ra\la \varphi_j|.
\ee
We expand smooth function $z=z(x)$ and $\psi=\psi(x)$ in the Fourier series
\bea
z(x) &=& \sum_{j=1}^\infty z^j\varphi_j(x),
\\
\psi(x) &=& \sum_{j=1}^\infty \psi_j\varphi_j(x).
\eea 
Then $z^{j}$, $\bar z^{j}$, are the local holomorphic coordinates 
and $\psi_{j}, \bar \psi_{j}$, are the components of a one-form.

Notice that the operator $H$ acts by multiplication by $H_j$ on each
subspace $V_j$. We identify the constant $H_j$ with the Einstein-K\"ahler
magnetic field on $M_j$ 
and consider the Hamiltonian system (\ref{544iga})
on the
state manifold $\cM$ with the Hamiltonian (\ref{542iga});
we also identify the parameter $q$ with the Planck constant,
\be
q=\frac{1}{\hbar}.
\ee
%\be
%\cH=\sum_{j=1}^\infty \cH_j,
%\ee
In the case of the complex hyperbolic space $M_j=\CC\HH^{d_j}$,
the Hamiltonian system has the form (\ref{5123iga}).
We postulate the dynamics of a quantum system to be determined by this
Hamiltonian flow with the initial conditions
\be
z^{j}(0)=0, \qquad
\psi_{j}(0)=\xi_{j}.
\ee
Thus, the quantum dynamics is given by
\bea
z(x) &=& \sum_{j=1}^\infty z^j(t)\varphi_j(x),
\\
\psi(x) &=& \sum_{j=1}^\infty \psi_j(t)\varphi_j(x).
\eea 
The $L^2$ norm of the one-form $\psi$ is
\be
|\psi|^2=\sum_{j=1}^\infty |\psi_j(t)|^2.
\ee
The geometric norm is conserved
\be
||\psi(t)||^2=4|\xi|^2=4\sum_{j=1}^\infty |\xi_j|^2
\ee

We define the parameters
\be
k_j=4\hbar\frac{{}|\xi_j|}{H_jb_j}.
\ee
%where $|\xi_j|^2=\bar\xi_{j}\xi_{j}$.
Since all subspaces fully decouple, the dynamics 
in each subspace is independent and depends crucially on the parameters
$k_j$. 
For $k_j<1$ we have the quantum regime (\ref{5132iga})
\bea
z^j(t) &=& 
4\hbar\frac{1}{H_j}
\frac{
\left\{i\left[1-\cos(2\omega_j t)\right]
+\sqrt{1-k_j^2}\sin(2\omega_j t)
\right\}}{2-k_j^2-k_j^2\cos(2\omega_j t)}
\bar \xi_j,
\label{5131igax}
\\
\psi_j(t) &=&
\frac{1}{2}\left\{
-\frac{k_j^2}{1-k_j^2}+\frac{2-k_j^2}{1-k_j^2}\cos(2\omega_j t)
-2i\frac{1}{\sqrt{1-k_j^2}}\sin(2\omega_j t)\right\}
\xi_j,
\nn\\
\label{5132igax}
\eea
where
\be
\omega_j=\frac{1}{2\hbar}H_j\sqrt{1-k_j^2}
=\sqrt{\frac{H_j^2}{4\hbar^2}-4\frac{|\xi_j|^2}{b_j^2}}
\ee
Of course, for $k_j=0$
we get the usual Schr\"odinger
dynamics
\be
\psi_j(t) =
\exp\left(-i \frac{H_j}{\hbar}t\right)
\xi_j
\label{5132igaxy}
\ee
The corrections to this behavior for small $k_j$ are given by (\ref{5138iga})
\be
\psi_j(t) = \left\{1+\frac{k_j^2}{2}\left[1-\exp(2i\omega_j t)\right]
+O\left(k_j^4\right)\right\}\exp(-2i\omega_j t)\xi_j,
\ee

%=============

For $k_j>1$ we have the classical regime (\ref{5144iga})
\bea
z^j(t) &=& 
4\hbar\frac{1}{H_j}
\frac{\left\{
i\left[\cosh(2\mu_j t)-1\right]
+\sqrt{k_j^2-1}\sinh(2\mu_j t)
\right\}}{k_j^2\cosh(2\mu_j t)+k_j^2-2}
\bar \xi_j,
\label{5143igax}\\
\psi_j(t) &=&
\frac{1}{2}\left\{
\frac{k_j^2}{k_j^2-1}+\frac{k_j^2-2}{k_j^2-1}\cosh(2\mu_j t)
-2i\frac{1}{\sqrt{k_j^2-1}}\sinh(2\mu_j t)\right\}
\xi_j,
\nn\\
\label{5144igax}
\eea
where
\be
\mu_j=\frac{1}{2\hbar }H_j\sqrt{k_j^2-1}
=\sqrt{4\frac{|\xi_j|^2}{b_j^2}-\frac{H_j^2}{4\hbar^2}}
\ee
Notice that the function $\psi_j(t)$ grows exponentially 
without bound as $t\to \infty$,
\be
\psi_j(t) \sim \frac{1}{4}\left(1-\frac{i}{\sqrt{k_j^2-1}}\right)^2
\exp(2\mu_j t)\xi_j.
\ee 
The quantum corrections to the classical behavior
as $k\to \infty$ are 
\bea
\psi_j(t) &=& 
\left\{
1-\frac{2i}{k_j}\tanh(\mu_j t)+O\left(\frac{1}{k^3}\right)
\right\}
\cosh^2\left(\mu_j t\right)\xi_j{}.
\eea
In the limit $k_j\to \infty$ they are
the geodesics
on the complex hyperbolic space $\CC\HH^1$,
\bea
z^j(t) &=& b_j\tanh\left(2\frac{|\xi_j{}|}{b_j} t\right)\frac{\bar 
\xi_j{}}{|\xi_j{}|},
\\
\psi_j(t) &=& 
\cosh^2\left(2\frac{|\xi_j{}|}{b_j} t\right)\xi_j{}.
\eea

Thus, the curvature, that is, the parameters $b_j$, provides an interpolation
between the quantum and the classical regime. 
The quantum world becomes classical at the bifurcation point
when periodic solutions become not periodic.

Notice that different modes $\psi_j$ can have different behavior
depending on the initial condition $\xi_j$. If a mode becomes
classical then it becomes the dominant one. For example, suppose that
$\mu_{N}$ is the largest one, that is, for any $j\ne N$,
\be
\mu_N > \mu_j.
\ee
Then the function $\psi(t)$ collapses to
\be
\psi(t,x)\sim \psi_{N}(t)\varphi_N(x)
\sim \frac{1}{4}\left(1-\frac{i}{\sqrt{1-k_N^2}}\right)^2
\exp(2\mu_N t)\xi_N \varphi_N(x)
\ee
with the characteristic time
\be
\tau_c=\max_{j\ne N}\frac{1}{|\mu_N-\mu_j|}.
\ee

Our main idea is that {\it the curvature of the hyperbolic space is determined
by gravity}, that is, by the mass/energy of the system, so that a macroscopic system
does not exhibit quantum behavior. Therefore,
we propose that the curvature is proportional to the energy, or
\be
b_j=4\hbar\frac{M}{m}|\xi|\frac{1}{ H_j}
\ee
where
\be
|\xi|^2=\sum_{j=1}^\infty|\xi_j|^2,
\ee
$m$ is the mass of the system and $M$ is some fundamental mass.
Then
\be
k_j=\frac{m}{M} \frac{|\xi_j|}{|\xi|}
\ee
For example, if we choose $M$ to be the Planckian mass
then
\be
k_j=\sqrt{\frac{Gm^2}{\hbar c}}\frac{|\xi_j|}{|\xi|}.
\ee
Then the parameter $\mu_j$ is
\be
\mu_j=\frac{H_j}{2\hbar}\sqrt{\frac{m^2}{M^2}\frac{|\xi_j|^2}{|\xi|^2}-1}.
\ee
Notice that 
\be
\frac{|\xi_j|}{|\xi|}\le 1.
\ee
Therefore, if $m<M$ then all modes are in the quantum regime with the unitary dynamics. 
There is no collapse.
However, if $m>M$ then the modes with the initial condition 
\be
|\xi_j|^2>\frac{M^2}{m^2-M^2}\sum_{i\ne j}|\xi_i|^2
\ee
will be in the classical regime,
with a collapse. 

%=============================
%{\bf 
%STOP 12/2/2024 
%}
%===========================================

%\section{Observables}
%\setcounter{equation}{0} 

Every physical observable is described by a 
tensor of type $(1,1)$ that can be identified with
a self-adjoint operator on the cotangent space
\be
A: T^*_{z}\cM\to T^*_{z}\cM,
\ee
such that for any $\varphi,\psi\in T^*_zM$,
\be
(A\varphi, \psi)=(\varphi, A\psi).
\ee
The expectation value of the measurement of the observable $A$
in a state $(z,\psi)$ is
\be
\left<A\right>_\psi=\frac{(\psi,A\psi)}{||\psi||^2},
\ee

We assign operators to the observables as follows. We assume that
all observables are described by the same operators as in the
standard Quantum Mechanics at the origin $z=0$. 
We extend them to the whole manifold by the parallel transport
along the dynamical trajectories, that is, 
\be
\frac{D A_\mu{}^\nu}{dt} = \dot z^\alpha \nabla_{\alpha}A_\mu{}^\nu
+\dot z^{\bar\alpha}\nabla_{\bar\alpha}A_\mu{}^\nu
=0
\ee
Since the connection is compatible with the metric,
it is easy to see that the self-adjointness condition
\be
\overline{A_\mu{}^\nu}
= g^{\bar\nu\alpha} A_{\alpha}{}^{\beta} g_{\beta \bar\mu}.
\ee 
is preserved under the parallel transport.
That is, if the operator $A$ is self-adjoint at the initial point $z=0$
then it remains self-adjoint along the dynamical trajectory.

%========================================================
%========================================================
%\section{Appendix: Observables}
%========================================
%==============
In particular, if the metric is diagonal
\be
g_{\mu\bar\nu}=e^{h_\mu}\delta_{\mu\nu},
%\qquad
%g^{\mu\bar\nu}=e^{-h_\mu}\delta_{\mu\nu},
\ee
with real $h_\mu$,
then the equation of parallel transport is
\bea
\frac{dA_\mu{}^\nu}{dt}
- \phi_{\mu\nu} A_{\mu}{}^{\nu}
=0,
\quad (\mbox{no summation!})
\eea
where
\be
\phi_{\mu\nu}=\dot z^\alpha \partial_\alpha (h_{\mu}-h_\nu);
\ee
therefore,
\be
A_\mu{}^\nu(t)=P_{\mu\nu}(t)A_\mu{}^\nu(0),
\ee
where
\be
P_{\mu\nu}(t)=\exp\left\{\int_0^t d\tau 
\phi_{\mu\nu}(\tau)\right\}
\ee

%=====================
%=================================================================

\end{document}